\documentclass[12pt]{iopart}
\usepackage{hyperref}
\usepackage{iopams}
\usepackage{setstack}
\usepackage{graphicx}
\usepackage{subfig}
\begin{document}
	\title{Twist angle, Strain, Corrugation and Moire Unit Cell in Twisted Bi-layer Graphene }
	%\title{Moire pattern and its commensurability in Twisted Bi-layer Graphene}
	\author{Veer Pal$^1$ and Ajay$^2$}
	\address{ Department of Physics, Indian Institute of Technology Roorkee$^{1,2}$}
	\ead{veerpal@ph.iitr.ac.in$^1$ and ajay@ph.iitr.ac.in$^2$}
	
	\begin{abstract}
		Knowledge of the internal configuration of carbon atoms inside a moire unit cell of twisted bi-layer graphene (TBG) would enhance the accuracy of many-body quantum mechanical calculations related to TBG. This work put forward a comprehensive theoretical study of moire pattern in TBG, supported with computational analysis; which seek a mechanism to determine the internal configuration of carbon atoms inside a moire unit cell of TBG. This study first time establishes that all twist angles are commensurate twist angles which produce perfectly periodic commensurate moire patterns of TBG. It is also first time established that strain appearing in moire patterns of TBG can occur purely due to intrinsic reasons. Taking some insight from available experimental data related to twisted bi-layer graphene systems and conventional bi-layer graphene systems, a mathematical model is also presented for corrugation in TBG. Finally we present an universal algorithm to determine the internal configuration of carbon atoms inside a moire unit cell of TBG, which is first of its kind.
		
	\end{abstract}
	\section{Introduction}\label{sec1}	
 Graphene based layered materials with a relative twist between the layers possess  many interesting properties e.g., moire pattern in their lattice structure, flat band near Dirac point \cite{2021-Lisi,2020-Utama}, Van-Hove singularities near Fermi level \cite{2010-Guohong,2019-Yonglong,2019-Alexander}, unconventional superconductivity \cite{1-2018-March-Cao,2019-Matthew,2021-Myungchul}, correlated insulator behavior \cite{2-2018-March-Cao}, anomalous Hall effect at half filling \cite{2022-Chun-Chih}, ferromagnetism \cite{2019-Sharpe,2022-Jiang}, Hofstadter butterfly \cite{2020-Bartholomew,2022-Nadia} and many more. It is proposed \cite{1-2018-March-Cao,2021-Myungchul} that investigation of electronic correlations in twisted bi-layer graphene may be helpful in solving the mystery of high temperature superconductivity, as well as it can help to understand other electronic correlations-based effects.\\ 
 Knowledge of unit cell of any material is essential for reliable quantum mechanical calculations related to that material. There are many research papers \cite{2007-Santos,2010-Shallcross,2010-Mele,2013-Yasumasa,2014-Kazuyuki} which provide very good description of commensurate moire pattern and report the relation between commensurate moire period $L_c$ and commensurate twist angle $\theta_c$ in TBG; but that description is not enough for development of an universal mechanism to determine the internal configuration of carbon atoms inside a moire unit cell of TBG. According to current understanding of moire pattern of TBG, ideal moire patterns of TBG should be defect free but experimentally observed moire patterns of TBG show presence of defects such as strain and broken rotational symmetry \cite{2019-Alexander,2019-Yonglong,2019-Youngjoon,2019-Yuhang}.\\
 In this article we offer a comprehensive theoretical study of moire pattern in TBG supported with computational analysis, which enables us to devise an universal algorithm for determination of the internal configuration of carbon atoms inside a moire unit cell of TBG. This study establishes that all moire patterns of TBG are perfectly periodic commensurate moire patterns. It is also established that strain appearing in moire patterns of TBG can occur purely due to intrinsic reasons. Taking some insight from available experimental data related to twisted bi-layer graphene systems and conventional bi-layer graphene systems, a mathematical model is developed for corrugation in TBG. Finally we present an universal algorithm to determine the internal configuration of carbon atoms inside a moire unit cell of TBG, which is first of its kind.\\
 The content of following sections in this article is organized as follows. In Sec. \ref{sec2}, we comprehensively discuss evolution of moire pattern of TBG to obtain relation of moire period with corresponding twist angle and finally reach at the conclusion that all twist angles are commensurate twist angles which produce perfectly periodic commensurate moire patterns of TBG. In Sec. \ref{sec3}, we discuss intrinsic reasons behind appearance of strain and rotational symmetry breaking in moire pattern of TBG. In Sec. \ref{sec4}, we present mathematical formulation for corrugation in structure of TBG. In Sec. \ref{sec5}, we present an universal algorithm for determination of internal configuration of carbon atoms inside a moire unit cell of TBG. In Sec. \ref{sec6}, we make conclusion.
 
	\section{Evolution of moire pattern of TBG}\label{sec2}
	
	\begin{figure} [htp]
		\centering
		\subfloat[]{\includegraphics[width=0.49\linewidth]{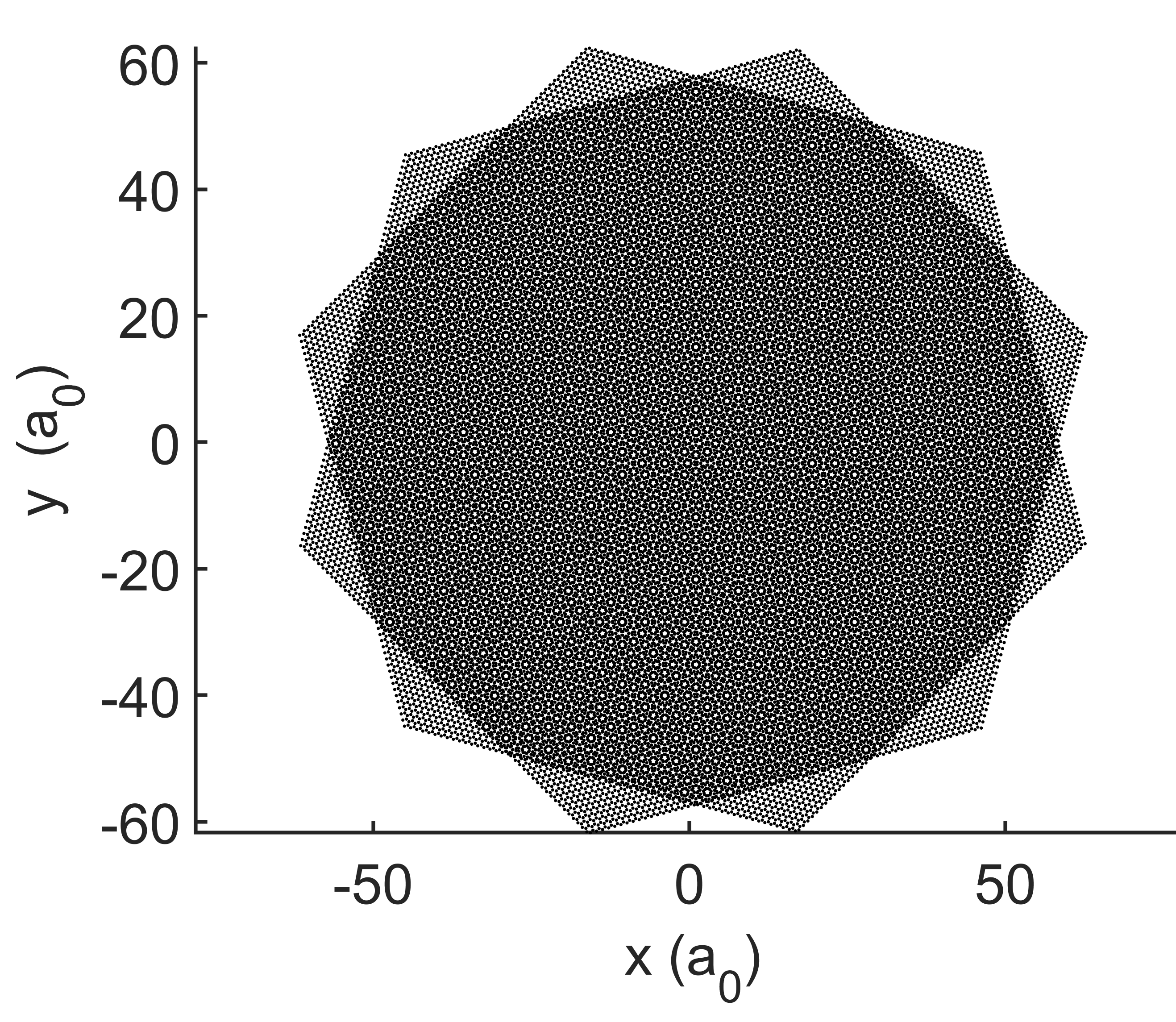}\label{fig1a}}
		\subfloat[]{\includegraphics[width=0.49\linewidth]{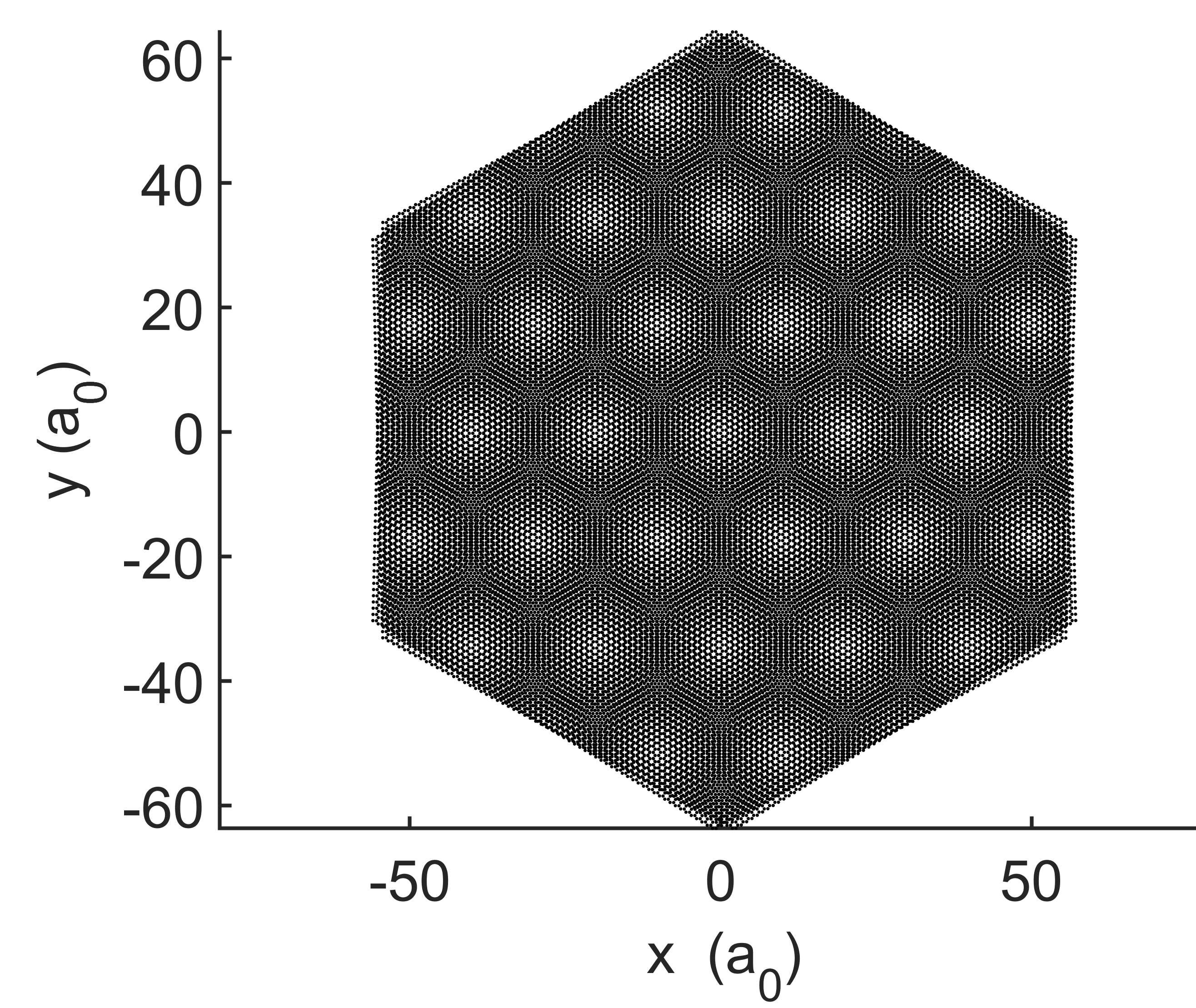}\label{fig1b}}
		\caption{(a) Incommensurate moire pattern of TBG  for twist angle equal to $30^{\circ}$, (b) commensurate moire pattern of TBG  for twist angle equal to $2.876^{\circ}$}	
	\end{figure}
    To understand the evolution of moire pattern of TBG, we start with basic understanding of TBG and following some reasonable considerations we reach to final conclusion.\\
    
    $\bullet$ Observed moire patterns in TBG systems are categorized in two categories: incommensurate moire patterns  (see figure \ref{fig1a}) and commensurate moire patterns (see figure \ref{fig1b}). Observed commensurate moire patterns of TBG show triangular lattice arrangement of moire lattice points (see Fig. 1(d,e,f) from ref.\cite{2019-Alexander}; Fig. 1(c) from ref. \cite{2019-Yonglong}; Fig. 1(g) from ref. \cite{2019-Youngjoon} ) with some strain and broken rotational symmetry \cite{2019-Yuhang}. Incommensurate moire patterns of TBG  don't seem to possess any periodicity.\\
    
    $\bullet$ To obtain the figures of moire pattern of TBG presented in this paper, we made computational code to simulate planar position of carbon atoms in a crystal of TBG. This code first simulates planar position of carbon atoms in a crystal of AA-stacked bi-layer graphene and then moire pattern of TBG is obtained after changing the planar position of carbon atoms of AA-stacked bilayer graphene according to rule of planar rotational transformation. To obtain TBG from AA-stacked bilayer graphene, we rotate the two graphene layers in opposite directions by half of the twist angle.\\
    
    $\bullet$ The lattice structure of moire pattern in TBG is considered to be modified form of conventional bi-layer graphene (either AA-stacked or AB-stacked bi-layer graphene); which is obtained after relative rotation, between two graphene layers of conventional bi-layer graphene, about an axis normal to their plane.\\
    
	$\bullet$ Conventional bi-layer graphenes are composed of graphene. In Graphene carbon atoms are arranged on a honeycomb structure made from hexagons, which are $sp^2$ hybridized, hence forming a planar structure. The distance between nearest neighbour carbon atoms (C-C bond length) is $a\approx1.415$ \AA. Lattice structure of graphene \cite{2009-Neto} can be considered to possess two inter-penetrating triangular lattices denoted as $A$ and $B$ sub-lattices, and each sub-lattice point is occupied by one carbon atom. The two sub-lattices in graphene are described by same Bravais lattice vectors  $\bi{a}_1=\frac{a_o}{2}(\sqrt{3}\hat{x}+\hat{y})$  and $\bi{a}_2=\frac{a_o}{2}(\sqrt{3}\hat{x}-\hat{y})$, but the translation vectors to neighbors are different in both sub-lattices. Angle between $\bi{a}_1$ and $\bi{a}_2$ is $60^\circ$. For same triangular lattice, one can also choose basis vectors which have $120^\circ$ angle between them. Here $a_o=\sqrt{3}a$ ($\approx2.46 $\AA) is the lattice constant for graphene. $\bi{b}_1=\frac{2\pi}{a_o\sqrt{3}}(\hat{k}_x+\sqrt{3}\hat{k}_y)$ and $\bi{b}_2=\frac{2\pi}{a_o\sqrt{3}}(\hat{k}_x-\sqrt{3}\hat{k}_y)$ are corresponding reciprocal lattice basis vectors, which cause hexagonal first Brillouin zone . The corners of first Brillouin zone of graphene are known as Dirac point \cite{2020-Liu} because at these points the 2D band structure of graphene shows the linear intersection of valence and conduction bands. According to our choice of basis vectors $\bi{a}_1$ and $\bi{a}_2$,  the coordinates of six Dirac points will be $\frac{4\pi}{3a_o} \left(\frac{\sqrt{3}}{2},\frac{-1}{2}\right)$, $\frac{4\pi}{3a_o} \left(\frac{\sqrt{3}}{2},\frac{1}{2}\right)$, $\frac{4\pi}{3a_o} \left(0,1\right)$, $\frac{4\pi}{3a_o} \left(\frac{-\sqrt{3}}{2},\frac{1}{2}\right)$, $\frac{4\pi}{3a_o} \left(\frac{-\sqrt{3}}{2},\frac{-1}{2}\right)$ and $\frac{4\pi}{3a_o} \left(0,-1\right)$. Dirac points of graphene are important even for twisted bilayer graphene because flat band in magic angle twisted bilayer graphene lies near Dirac point of graphene \cite{2020-Utama}.\\
	
	$\bullet$ We try to see that as we move from conventional bi-layer graphene to twisted bi-layer graphene through relative rotation between the graphene layers of conventional bi-layer graphene, how the z component of position of carbon atoms changes with change in the planar component. Initially, in the conventional bi-layer graphene the z component of position of carbon atoms is found to be same in whole graphene layer. The z component of position of carbon atoms changes after change in the plane component due to relative rotation between the graphene layers. First, we try to study the plane component of the position of carbon atoms in TBG and then the z component is decided according to plane component of position.  Although our approach is different, but our results agree with the results obtained through vdW-DFT and LDA-DFT simulations reported in research paper associated with reference \cite{2014-Kazuyuki}. In further discussion of this section planar components of position of carbon atoms are considered and the z component is decided in section \ref{sec4}.\\
		
	$\bullet$ The conventional bi-layer graphene has two graphene layers, one layer stacked on top of another layer. Let $A_1$, $B_1$ denote sub-lattices of lower layer and $A_2$, $B_2$ denote sub-lattices of upper layer. In AA-stacked bi-layer graphene $A_1$ sub-lattice coincides with $A_2$ sub-lattice and $B_1$ sub-lattice coincide with $B_2$ sub-lattice. In AB-stacked bi-layer graphene $A_1$ sub-lattice coincides with $B_2$ sub-lattice but $B_1$ sub-lattice do not coincide with $A_2$ sub-lattice. In both the cases, lattice points of at least one sub-lattice of lower graphene layer coincides with lattice points of one sub-lattice of upper graphene layer, let such lattice points be named as coinciding lattice points of conventional bi-layer graphene.\\
	
	$\bullet$ The origin can be considered to be situated at one of the coinciding lattice points of conventional bi-layer graphene, other choices are also possible for origin but this choice simplifies the description and produces most appropriate results. Starting from conventional bi-layer graphene, when a graphene layer is rotated ( twisted) with respect to other graphene layer about the z-axis passing through the origin, a moire pattern is generated in the lattice structure of bi-layer graphene which evolves with twist angle. \\
	
	$\bullet$ One moire lattice point of TBG can be considered at origin where lattice points of two  graphene layers coincide; this choice simplifies the identification of moire lattice point. Since we are considering one moire lattice point to be situated at origin; therefore, only those positions of moire pattern can be regarded as moire lattice point where lattice points of two graphene layers coincide and the combination of coinciding lattice points  is of same type as it is at origin, i.e., If we consider AA-stacked bi-layer graphene initially, then at position of moire lattice points, lattice points of $A_1$ and $A_2$ (or $B_1$ and $B_2$) sub-lattices will coincide and If we consider AB-stacked bi-layer graphene initially, then at position of moire lattice points, lattice points of $A_1$ and $B_2$ sub-lattices will coincide.\\
	
	$\bullet$  Since whole graphene layer is being rotated, therefore, even after planar rotational transformation, the lattice of rotated graphene layer will remain perfectly periodic triangular lattice. Since the lattice of individual graphene layers of TBG retains its periodicity, the moire pattern of TBG will be result of combination of two perfectly periodic planar lattices, therefore, it will be perfectly periodic.  Due to perfect periodicity of moire pattern of TBG; for a particular twist angle, either there will be no moire lattice point other than origin, or if any other moire lattice point exists then the configuration of carbon atoms around each moire lattice point will be same as it is around origin. \\
	
	$\bullet$ Due to three-fold rotational symmetry of graphene layer for the rotation about the normal axis passing through any carbon atom, no moire pattern will be generated for twist angles equal to integer multiples of $120^\circ$. A relative rotation of  $60^\circ$  between two graphene layers of AA-stacked bi-layer graphene will convert  it into a structure that will have appearance similar to AB stacked bi-layer graphene or vice versa; this situation can be considered as the creation of no moire pattern or creation of a moire pattern with infinitely large moire period.  The moire pattern with same characteristics will be generated for twist angles $\theta$, $-\theta$, $60^\circ+\theta$ and $60^\circ-\theta$. The moire patterns generated for twist angles $\theta$, $-\theta$, $60^\circ+\theta$ and $60^\circ-\theta$ will be practically indistinguishable. The moire pattern of TBG will possess 6-fold rotational symmetry. Due to 6-fold rotational symmetry of moire pattern of TBG, it is sufficient to investigate the commensurate twist angles lying between $0^\circ$ and $30^\circ$. In further discussion we shall consider twist angle which lie in the range $\left(0^{\circ},30^{\circ}\right)$.\\
	
	$\bullet$  A perfectly periodic moire pattern with 6-fold rotational symmetry will have triangular lattice arrangement of moire lattice points. Such moire patterns of TBG are known as commensurate moire patterns and the underlying twist angle as commensurate twist angle.\\ 
	
	$\bullet$  All coinciding lattice points of conventional bi-layer graphene have same type of combination of lattice points. Before relative rotation between the graphene layers of conventional bi-layer graphene all coinciding lattice points of upper graphene layer and lower graphene layer coincide. After relative rotation between the graphene layers of conventional bi-layer graphene by a commensurate twist angle, the number of generated coinciding lattice points in TBG will be much less as compared to conventional bi-layer graphene. Moire lattice points of commensurate TBG will be those coinciding lattice points which will be generated after relative rotation between the layers of coinciding lattice points of conventional bi-layer graphene. Commensurate twist angles will be those which generate a triangular lattice pattern of moire lattice points after relative rotation between the layers of coinciding lattice points of conventional bi-layer graphene. The evolution of moire pattern of TBG can be studied by studying of movement of coinciding lattice points of conventional bi-layer graphene under planar rotational transformation.\\ 
	
	\begin{figure} [htp]
		\centering
		\subfloat[]{\includegraphics[width=0.49\linewidth]{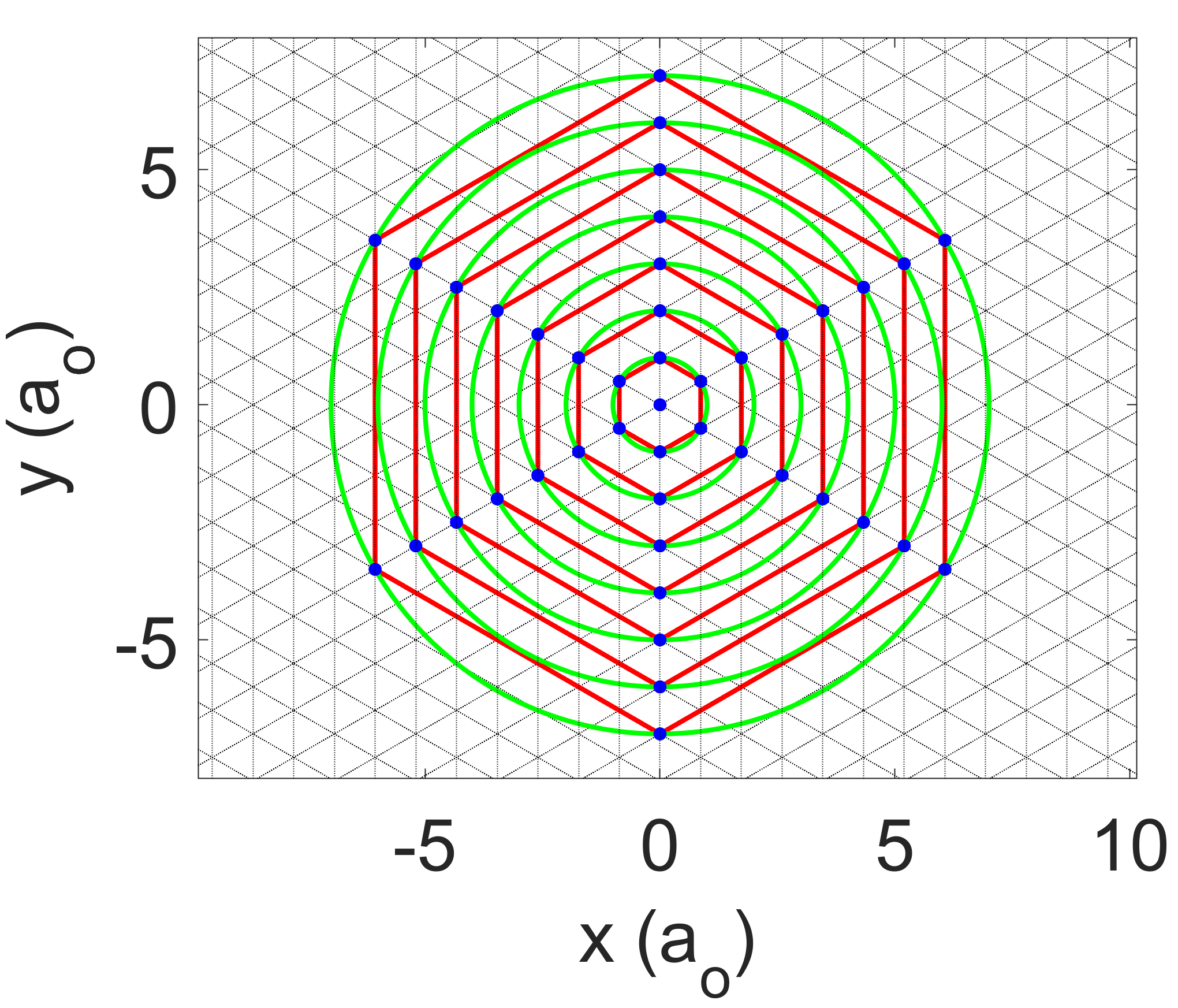}\label{fig2a}}
		\subfloat[]{\includegraphics[width=0.49\linewidth]{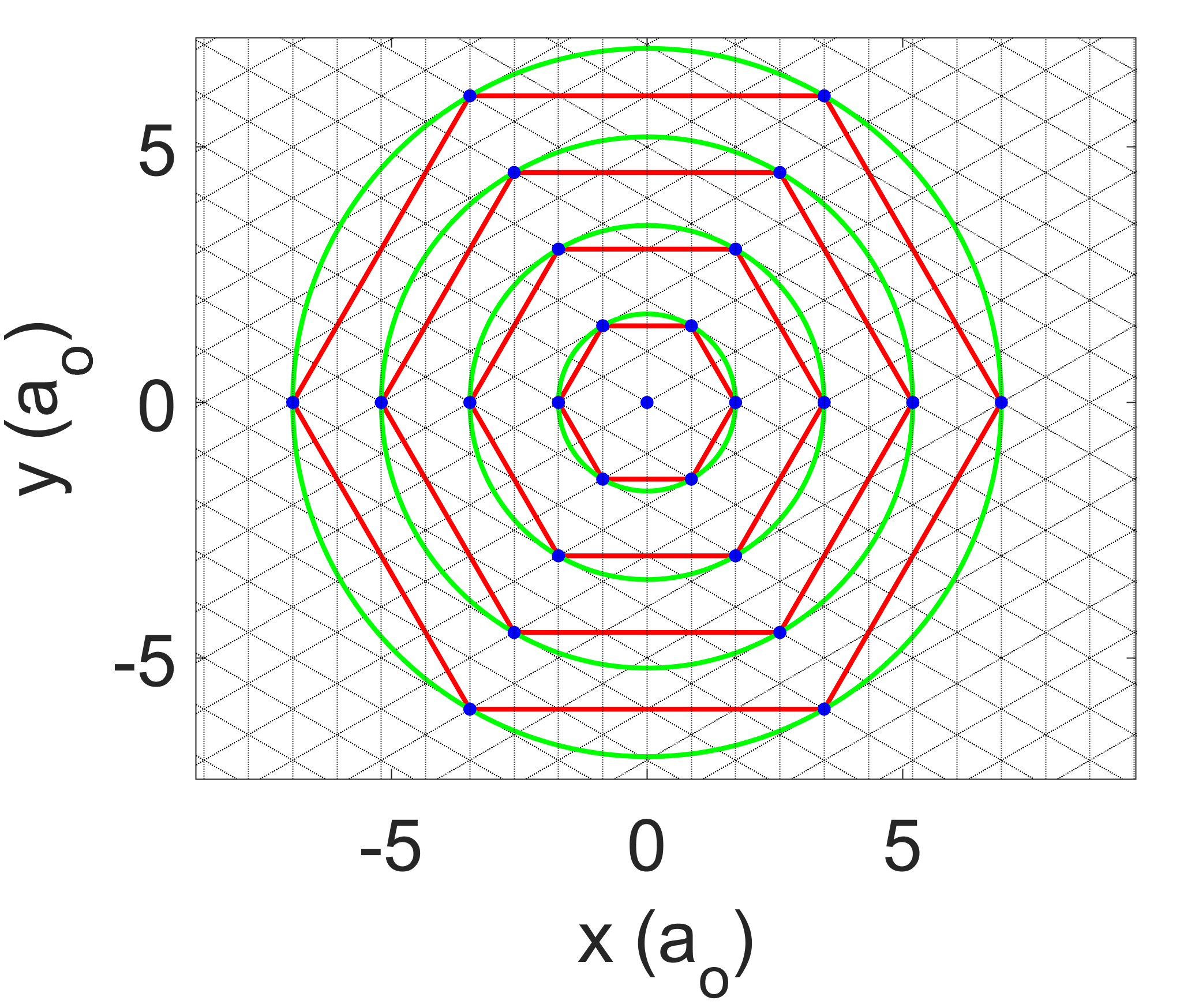}\label{fig2b}}\\
		\caption{(a) Some sets of triangular lattice points which are equidistant from origin and have distance from origin  equal to integer multiple of lattice constant. (b) Some sets of triangular lattice points which are equidistant from origin and have distance from origin equal to integer multiple of $\sqrt{3}$ times of lattice constant.}	
	\end{figure}
	
	$\bullet$ From the geometry of triangular lattice it can be easily understood that in a triangular lattice of moire lattice points the set of moire lattice points lying nearest to origin will have only 6 moire lattice points,  which will be equidistant from origin and form a regular hexagon. The side length of this hexagon will be equal to moire period $L_c$ of moire pattern. Similar type of arrangement of moire lattice points will exist around each moire lattice point. This is also to note that the sets of those moire lattice points which are equidistant from origin and  have their distance from origin as integer multiple of $L_c$ or integer multiple of $\sqrt{3}L_c$ also contain only 6 moire lattice points (most of the times) and they  also form a regular hexagon which have center at origin (see figure \ref{fig2a},\ref{fig2b}).
	
    $\bullet$ To find out commensurate twist angles and corresponding commensurate moire period in TBG, it will be sufficient to inspect the creation of those sets of  moire lattice points , which contain only 6 moire lattice points equidistant from origin  and they form a regular hexagon having center at origin. The distance of these moire lattice points from the origin will be either integer multiple of $L_c$ or integer multiple of $\sqrt{3}L_c$. Many of those sets will generate to same value of  commensurate twist angle $\theta_c$ . Corresponding to any commensurate twist angle $\theta_c$, the commensurate moire period $L_c$  will be obtained from that set which has minimum value of side length of  hexagon of moire lattice points.\\
		
	\begin{figure} [htp]
		\centering
		\includegraphics[width=0.99\linewidth]{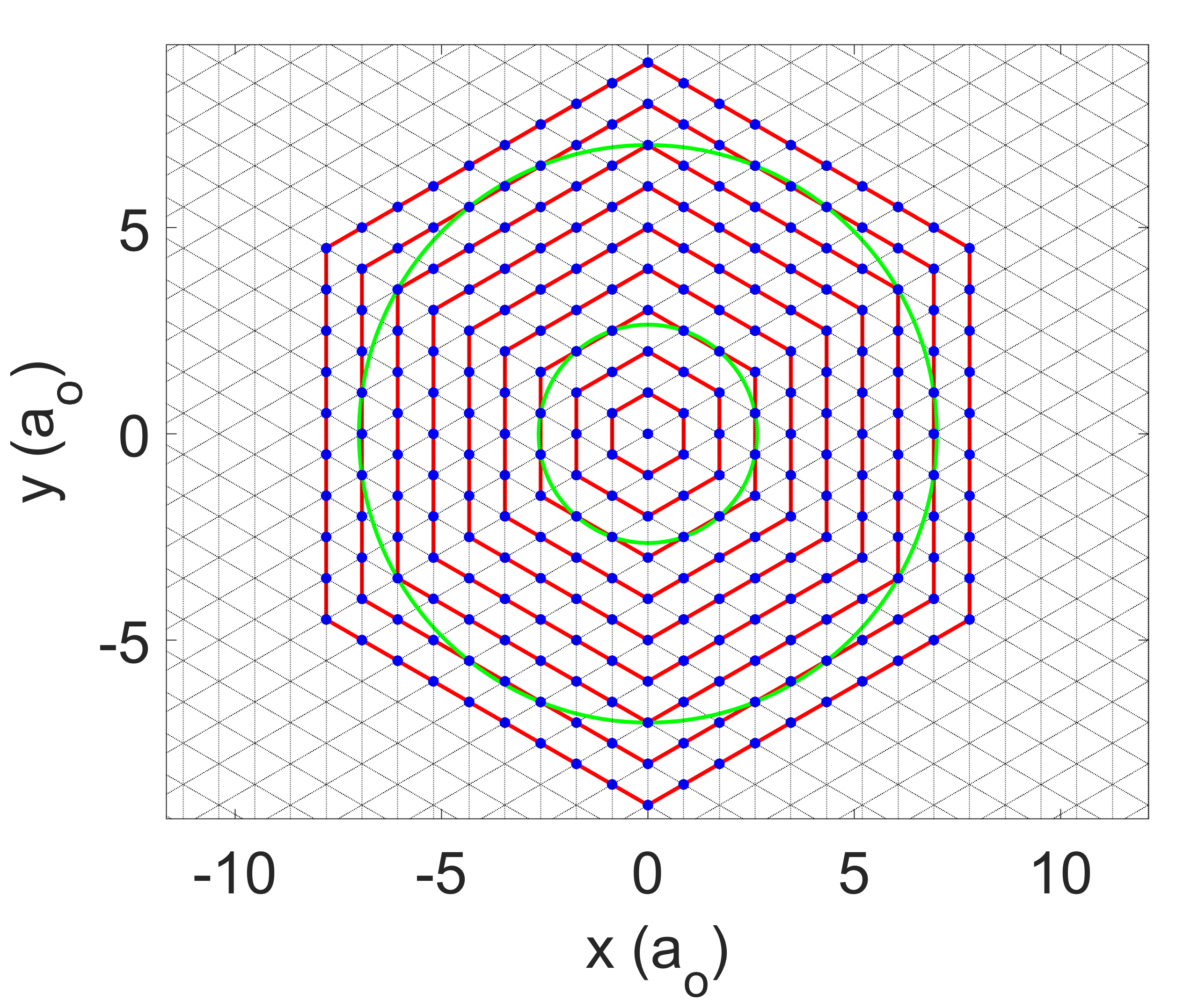}
		\caption{ Coinciding lattice points (triangular lattice points) arranged on hexagons, lattice points having same distance from origin fall on same circle.}\label{fig3}	
	\end{figure}
	
	$\bullet$ Coinciding lattice points of conventional bi-layer graphene form a triangular lattice ; therefore, they can be arranged on those concentric regular hexagons (see figure \ref{fig3}) which have center at origin and  have sides of length  $na_o$  ($n$ is a natural number) and the sides of those hexagons are either parallel to vectors $\pm\bi{a}_1$,$\pm\bi{a}_2$,$\pm \left(\bi{a}_1-\bi{a}_2\right)$ (if the angle between basis vectors is $60^\circ$) or  parallel to vectors $\pm\bi{a}_1$,$\pm\bi{a}_2$,$\pm \left(\bi{a}_1+\bi{a}_2\right)$ (if the angle between basis vectors is $120^\circ$). Spacing between any two consecutive hexagons will be $a_o\frac{\sqrt{3}}{2}$. The coinciding lattice points lying on a hexagon can be categorized in three categories; corner points, midpoint of sides, and points lying between corner and midpoint of sides.In further discussion these hexagons will be written as HEXAGON$^*$.\\
	
	$\bullet$ To simplify the description, we consider AA-stacked bi-layer graphene initially and origin at a site where $A_1$ sub-lattice coincide with $A_2$ sub-lattice. Before rotation of graphene layers  $A_1$ sub-lattice completely coincide with $A_2$ sub-lattice. The moire pattern depends on twist angle of one graphene layer relative to other graphene layer. For simplicity of description, here we consider that the lower graphene layer is static, and the upper graphene layer is rotated about the z axis passing through origin. \\
	
	$\bullet$ After rotation of upper layer  by a commensurate twist angle $\theta_c$ ($0^\circ <\theta_c<30^\circ$) about the z-axis passing through origin , not all but some of the lattice points of   $A_2$ sub-lattice will coincide with lattice points of $A_1$ sub-lattice, which will be moire lattice points (coinciding lattice point of TBG). Each moire lattice point  will be generated by transfer of a lattice point of $A_2$ sub-lattice from its initial position to that position which was previously occupied by a lattice point of $A_2$ sublattice equidistant from origin. \\
	
	$\bullet$ All the  lattice points of $A_1$ and $A_2$ sub-lattice having same distance from origin generally lie on same HEXAGON$^*$ (see lattice points on smaller circle of figure \ref{fig3}) but sometimes some of those lattice points might also lie on different HEXAGONS$^*$ (see lattice points on bigger circle of figure \ref{fig3}).\\ 
	
	$\bullet$ Those sets which contain only 6 moire lattice points equidistant from origin,which form a regular hexagon  having center at origin; can be generated  by only those planar rotational transformations, which occur between those lattice points of $A_2$ sub-lattice which are equidistant from origin and lie between corners and midpoint of sides of same HEXAGON$^*$. If we consider the planar rotational transformation between those lattice points of $A_2$ sub-lattice which lie on different HEXAGON$^*$ and have same distance from origin, then the set of  generated  moire lattice points equidistant from origin  will contain 12 moire lattice points. In other possibilities of commensurate planar rotational transformations, the commensurate twist angle will be integer multiple of $60^\circ$. \\
	
	$\bullet$ To detect commensurate twist angles of TBG, it is sufficient to study those planar rotational transformations, which occur between those lattice points of $A_2$ sub-lattice which are equidistant from origin and lie between corners and midpoint of sides of same HEXAGON$^*$.\\
	
	$\bullet$ Consider the lattice points of $A_2$ sub-lattice lying on a HEXAGON$^*$ of side length equal to $na_o$(In further discussion this HEXAGON$^*$ will be written as HEXAGON$^{*n}$) .
	Between a corner and mid-point of a side of this HEXAGON$^{*n}$ there will be $(p-1)$ lattice points of $A_2$ sub-lattice where $p=(\frac{n+rem(n,2)}{2})$  and $rem(n,2)=0$ if $n$ is even or $rem(n,2)=1$ if $n$ is odd. Let us assign an index $k=1,2,...(p-1)$ to these points. The distance of $k^{th}$ lattice points of $A_2$ sub-lattice from mid-point of its parent side will be equal to $ka_o-rem(n,2)\frac{a_o}{2}$. The distance of this  $k^{th}$ lattice points of $A_2$ sub-lattice from origin will be equal to $L_{n,k}=\sqrt{\left(na_o\frac{\sqrt{3}}{2}\right)^2+\left(ka_o-rem(n,2)\frac{a_o}{2}\right)^2}=\frac{a_o}{2}\sqrt{3n^2+\left(2k-rem(n,2)\right)^2}$.\\
	
	\begin{figure}[htp]
		\centering
		\includegraphics[width=0.99\linewidth, scale=1.0]{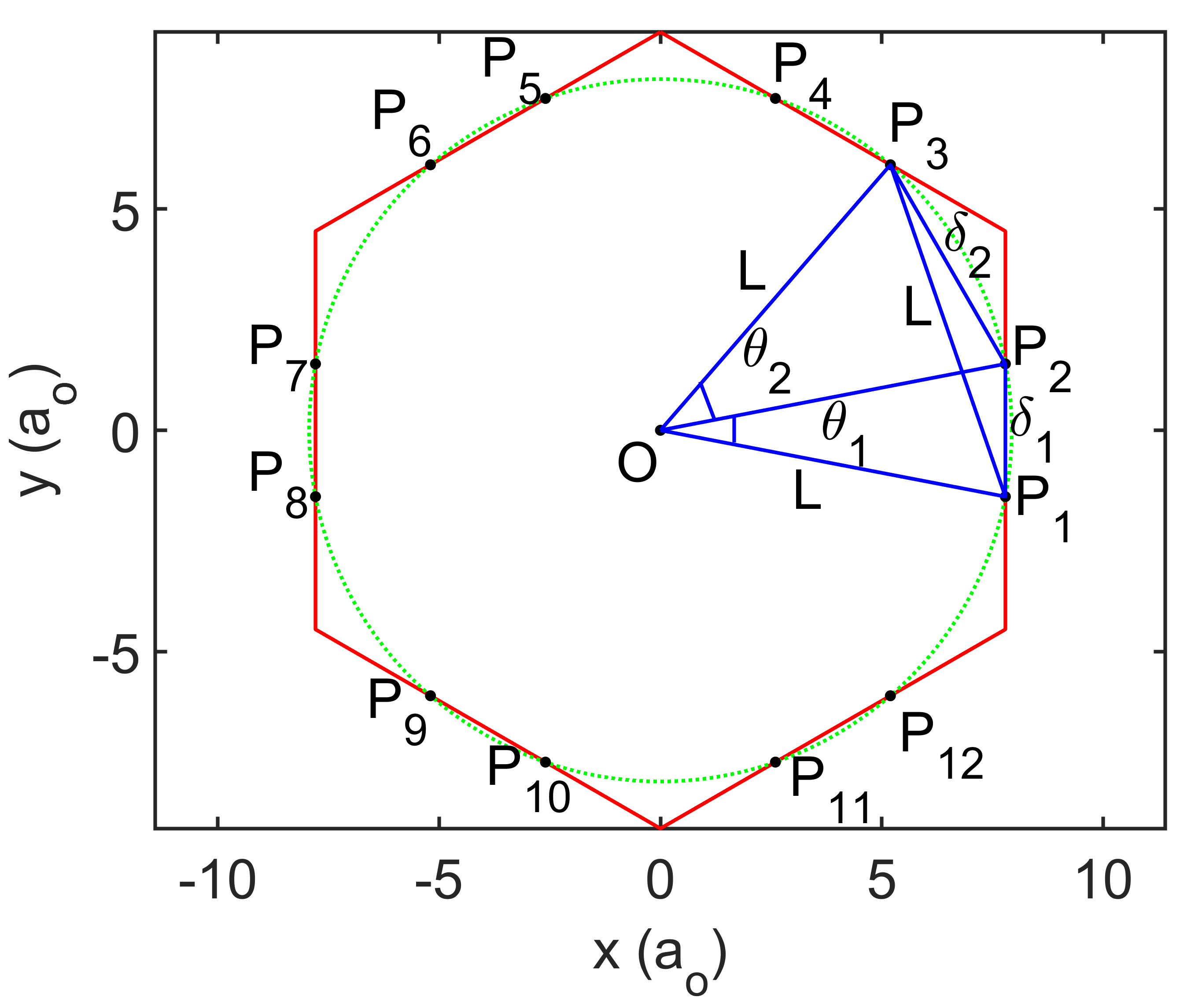}
		\caption{Commensurate rotation in TBG }	
		\label{fig4}
	\end{figure}

	$\bullet$ Corresponding to each index $k$ there will be 12  such lattice points of $A_2$ sub-lattice on a HEXAGON$^{*n}$  which will have same distance from origin (as shown in figure \ref{fig4}; $n,k$ part of index is omitted in figure); These can be denoted by $P_{1,n,k},P_{2,n,k},...,P_{12,n,k}$. The distance between any two consecutive points taken in cyclic order will be equal to $\delta_{1,n,k}=\left(2k-rem(n,2)\right)a_o$ if they are situated on same side of HEXAGON$^{*n}$ or it will be equal to $\delta_{2,n,k}=\sqrt{3}\left(n-2k+rem(n,2)\right)a_o$ if they are situated on adjacent sides of HEXAGON$^{*n}$. The distance between alternate points will be equal to $L_{n,k}$, i.e., distance of these points from origin; thus such points will make $60^\circ$ angle at origin.\\

	$\bullet$ To investigate the commensurate twist angles lying between $0^\circ$ and $30^\circ$, it is sufficient to investigate the commensurate planar rotational transformation among any 3 consecutive lattice points of $A_2$ sub-lattice lying on same HEXAGON$^{*n}$. Let us consider rotational transformation among 3 consecutive lattice points of $A_2$ sub-lattice denoted by $P_{1,n,k}, P_{2,n,k}, P_{2,n,k}$. For commensurate planar rotational transformation among these three lattice points there can be three commensurate displacements: $P_{1,n,k}\rightarrow P_{2,n,k}=\delta_{1,n,k}$, $P_{2,n,k}\rightarrow P_{3,n,k}=\delta_{2,n,k}$ and $P_{1,n,k}\rightarrow P_{3,n,k}=L_{n,k}$. The commensurate planar rotational transformation corresponding to commensurate displacement $P_{1,n,k}\rightarrow P_{3,n,k}=L_{n,k}$ will occur for twist angle equal to $60^\circ$, hence no moire pattern will be generated. The commensurate planar rotational transformation corresponding to commensurate displacement $P_{1,n,k}\rightarrow P_{2,n,k}=\delta_{1,n,k}$ will occur for twist angle equal to $\theta_{1,n,k}=2\sin^{-1}\left(\frac{\delta_{1,n,k}}{2L_{n,k}}\right)$ and the commensurate planar rotational transformation corresponding to commensurate displacement $P_{2,n,k}\rightarrow P_{3,n,k}=\delta_{2,n,k}$ will occur for twist angle equal to $\theta_{2,n,k}=2\sin^{-1}\left(\frac{\delta_{2,n,k}}{2L_{n,k}}\right)$. \\
	
	$\bullet$ It is worth noting that $\theta_{1,n,k}+\theta_{2,n,k}=60^\circ$; Therefore, moire pattern with same characteristics will be generated for twist angles $\theta_{1,n,k}$  and $\theta_{2,n,k}$. Therefore, corresponding to moire lattice point generated at distance $L_{n,k}$ from origin, only one of $\theta_{1,n,k}$ and $\theta_{2,n,k}$, whichever is smaller than $30^\circ$, is chosen as commensurate twist angle $\theta_c$, also the corresponding commensurate displacement $\delta_c$ is accordingly chosen from $\delta_{1,n,k}$ and $\delta_{2,n,k}$ .\\
	
	$\bullet$  All those pairs of $\left(n,k\right)$ which have same ratio of $\delta_{n,k}$  to $L_{n,k}$ produce same value of $\theta_c$ . Therefore, while choosing moire period ($L_c$) and minimum commensurate displacement ($\delta_c$) corresponding to any commensurate twist angle ($\theta_c$), we choose that pair of $\left(n,k\right)$ which gives the lowest value of ${L_{n,k}}$; the value of $\delta_{n,k}$ corresponding to this pair of $\left(n,k\right)$  is chosen as the value of minimum commensurate displacement ($\delta_c$);  and the value of $\L_{n,k}$ corresponding to this pair of $\left(n,k\right)$ is chosen as the value of moire period ($L_c$). \\
	
	$\bullet$ Using this understanding of commensurate moire patterns of TBG, we wrote a computational code \cite{TBG_L_vs_theta} in MATLAB to generate a list associating commensurate moire periods with corresponding commensurate twist angles and minimum commensurate displacement. Theoretically the upper limit of commensurate moire period of TBG ($Max.L_c$) can go up to infinity. For practical purposes first time we chose the upper limit of commensurate moire period of TBG to be equal to $100 a_o$ and second time we chose it to be equal to $300 a_o$. First time (for $Max.L_c=100 a_o$) the generated list \cite{100ao_TBG_Lc_vs_thetac} contained 1384 distinct twist angles lying between $0^{\circ}$ and $30^{\circ}$. Second time (for $Max.L_c=300 a_o$) the generated list \cite{300ao_TBG_Lc_vs_thetac} contained 12399 distinct twist angles lying between $0^{\circ}$ and $30^{\circ}$ (generated tables are too large to include in main article, therefore we have provided their online link along with link of MATLAB code). If we keep increasing the upper limit of commensurate moire period of TBG to further higher values then more and more commensurate twist angles appear in the generated list. This created a doubt about the total number of distinct commensurate twist angles for TBG, lying between $0^{\circ}$ and $30^{\circ}$.\\
	
	\begin{figure}[htp]
		\centering
		\includegraphics[width=0.99\linewidth, scale=1.0]{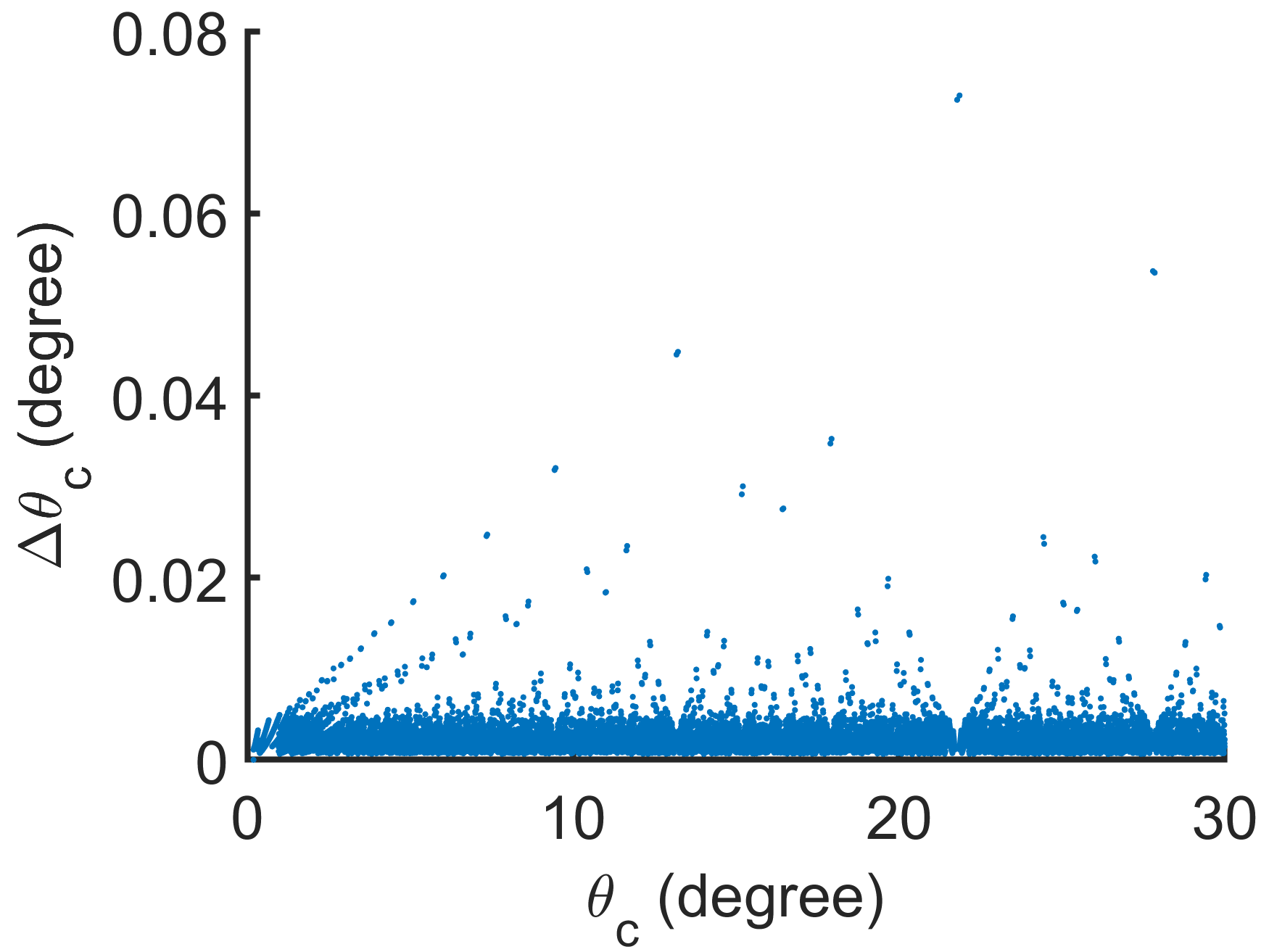}
		\caption{The gap between any two consecutive values of commensurate twist angles of TBG for $Max.L_c=300 a_o$}	
		\label{fig5}
	\end{figure}
	
	$\bullet$ To clear the doubt we arranged the twist angles of generated list \cite{300ao_TBG_Lc_vs_thetac} (for $Max.L_c=300 a_o$) in ascending order and observed that the twist angles of the list lie between $0.191766393304777^{\circ}$ and $29.9991841229813^{\circ}$. We computed the value of gap ($\Delta\theta_c$) between consecutive values of commensurate twist angles. Variation of gap with commensurate twist angle is shown in figure \ref{fig5}.  Average value of the gap  between any two consecutive values of commensurate twist angles  is $0.0024^{\circ}$ and the maximum value of gap between any two consecutive values of commensurate twist angles is  $0.0729^{\circ}$.\\
	
    $\bullet$ For many values of the chosen upper limit of commensurate moire period ($Max.L_c$); we computed the value of average gap ($Avg.\Delta\theta_c$)  and maximum gap ($Max.\Delta\theta_c$) between any two consecutive twist angles. We found that for large values of $Max.L_c$ the values of $Avg.\Delta\theta_c$, and $Max.\Delta\theta_c$ moves towards practically zero value. For example, when we choose $Max.L_c=300 a_o$ we get  $Avg.\Delta\theta_c=0.0024^{\circ}$ and$Max.\Delta\theta_c=0.0729^{\circ}$. For  $Max.L_c=1000 a_o$ we get  $Avg.\Delta\theta_c=0.00021722^{\circ}$ and$Max.\Delta\theta_c=0.0217^{\circ}$. For  $Max.L_c=5000 a_o$ we get  $Avg.\Delta\theta_c=0.0000087029^{\circ}$ and$Max.\Delta\theta_c=0.0043^{\circ}$.For  $Max.L_c=10000 a_o$ we get  $Avg.\Delta\theta_c=0.0000021762^{\circ}$ and$Max.\Delta\theta_c=0.0022^{\circ}$. We see that as we increase the value of  $Max.L_c$ to very large value, the values of  ($Avg.\Delta\theta_c$) and ($Max.\Delta\theta_c$) move towards practically zero value. Therefore, we can say that for infinitely large value of $Max.L_c$  the values of ($Avg.\Delta\theta_c$) and ($Max.\Delta\theta_c$) will become practically zero. Therefore, it can be inferred that all twist angles of TBG are commensurate twist angles which will generate perfectly periodic moire pattern.\\
     
    $\bullet$ The underlying reason behind this behaviour is explained as follows. Values of commensurate twist angles come from formula of $\theta_{1,n,k}$ ($\theta_c$ takes value either $\theta_{1,n,k}$ or $60^{\circ}-\theta_{1,n,k}$.), which depends on pair of integers $\left(n,k\right)$. The number of possible values of integer $k$ increases with value of $n$. for large values of $n$, we will get more pairs of integers $\left(n,k\right)$ . Some of those pairs of integers $\left(n,k\right)$ will produce those values of $\theta_{1,n,k}$ which were already produced for smaller value of $n$. But it is quite possible that some of those pairs of integers $\left(n,k\right)$ may produce new values of $\theta_{1,n,k}$ which were not produced for any smaller value of $n$. Increasing the upper limit of commensurate moire period ($Max.L_c$) increases upper limit of integer $n$ and new commensurate twist angles get detected which were undetected for smaller values of $Max.L_c$.  As the value of $Max.L_c$ increases the number of detected commensurate twist angles increase while the range of commensurate twist angles $\left(0^{\circ},30^{\circ} \right)$ remain same, therefore the gap between any two consecutive commensurate twist angles decreases.
    
	\section{Apparent strain and rotational symmetry breaking in moire pattern of commensurate TBG}\label{sec3}
	
	$\bullet$ For further investigation of commensurate moire pattern in TBG, we made computational code \cite{AA_to_TBG_Moire_pattern_strain_check}  to simulate planar position of carbon atoms in a crystal of TBG. TBG is obtained after introducing relative twist between two graphene layers of hexagonal crystal of conventional bi-layer graphene which has side length equal to $\frac{2}{\sqrt{3}}L_c$. Side of hexagonal crystal of conventional bi-layer graphene is chosen to be $\frac{2}{\sqrt{3}}L_c$, so that it can at least accommodate those six  moire lattice points which are generated nearest to origin.\\
	 
	$\bullet$ Initially, we consider AA-stacked bi-layer graphene and consider origin to be situated at a site where lattice points of $A_1$ and $A_2$ sub-lattice coincide. We rotate one graphene layer with respect to other graphene layer about the z-axis passing through the origin; moire pattern of TBG is obtained. One moire lattice point is considered to be situated  at origin, which lies in the center of an AA-type looking region of TBG. It is expected that locations of other moire lattice points will be in the center of an AA-type looking region of TBG. We understand the centers of AA-type looking regions of TBG as the apparent moire lattice points and the distance between two apparent moire lattice points as the apparent moire period.\\
	
	$\bullet$ We tried to simulate moire patterns of TBG for many commensurate twist angles near magic angle and each time the apparent moire lattice points appeared to have triangular symmetry, which was as expected; but sometimes the value of apparent moire period appeared to be quite different from the value of moire period ($L_c$),  supplied in the code, which was unexpected. \\ 
	
	\begin{figure} [htp]
		\centering
		\subfloat[]{\includegraphics[width=0.49\linewidth]{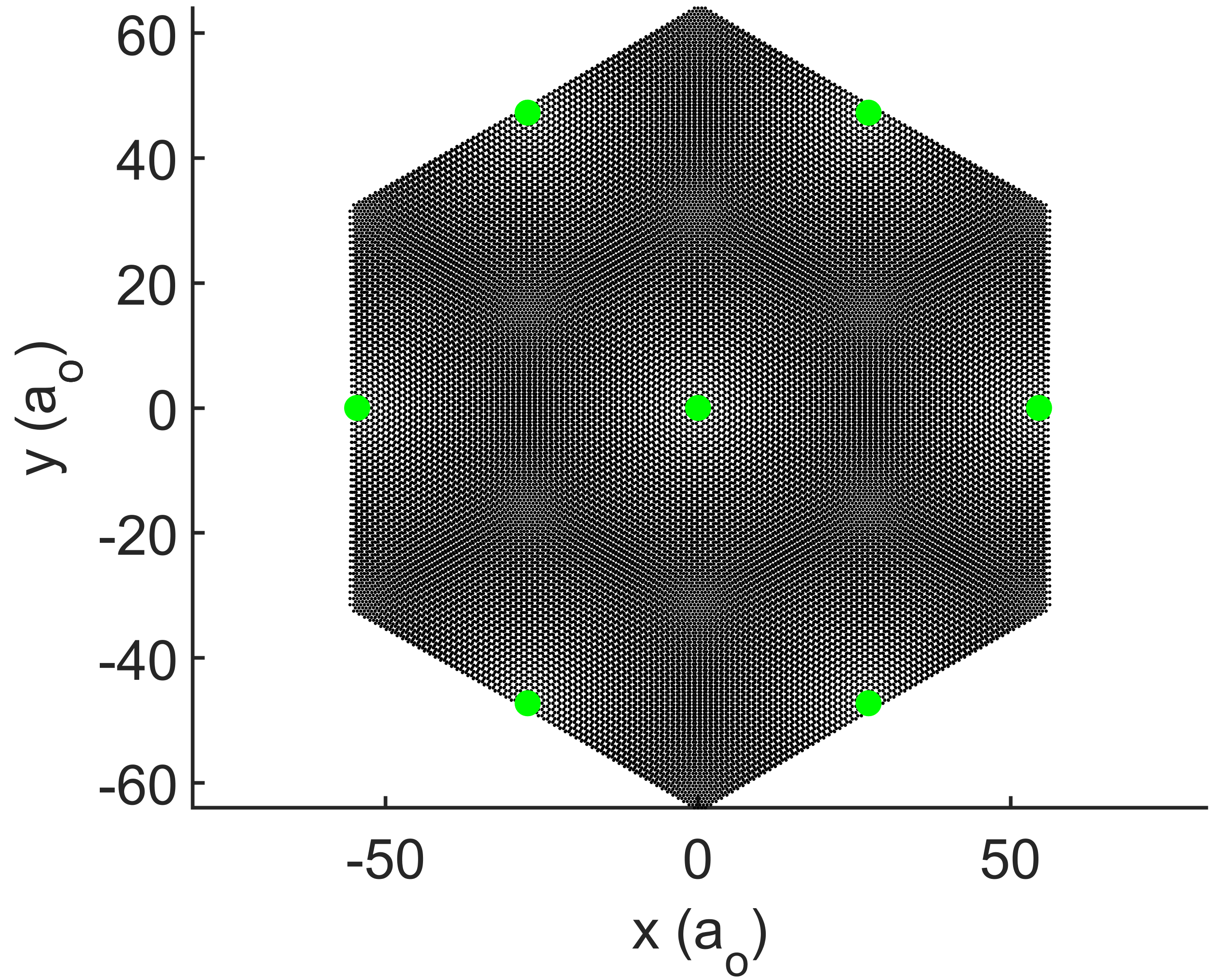}\label{fig6a}}
		\subfloat[]{\includegraphics[width=0.49\linewidth]{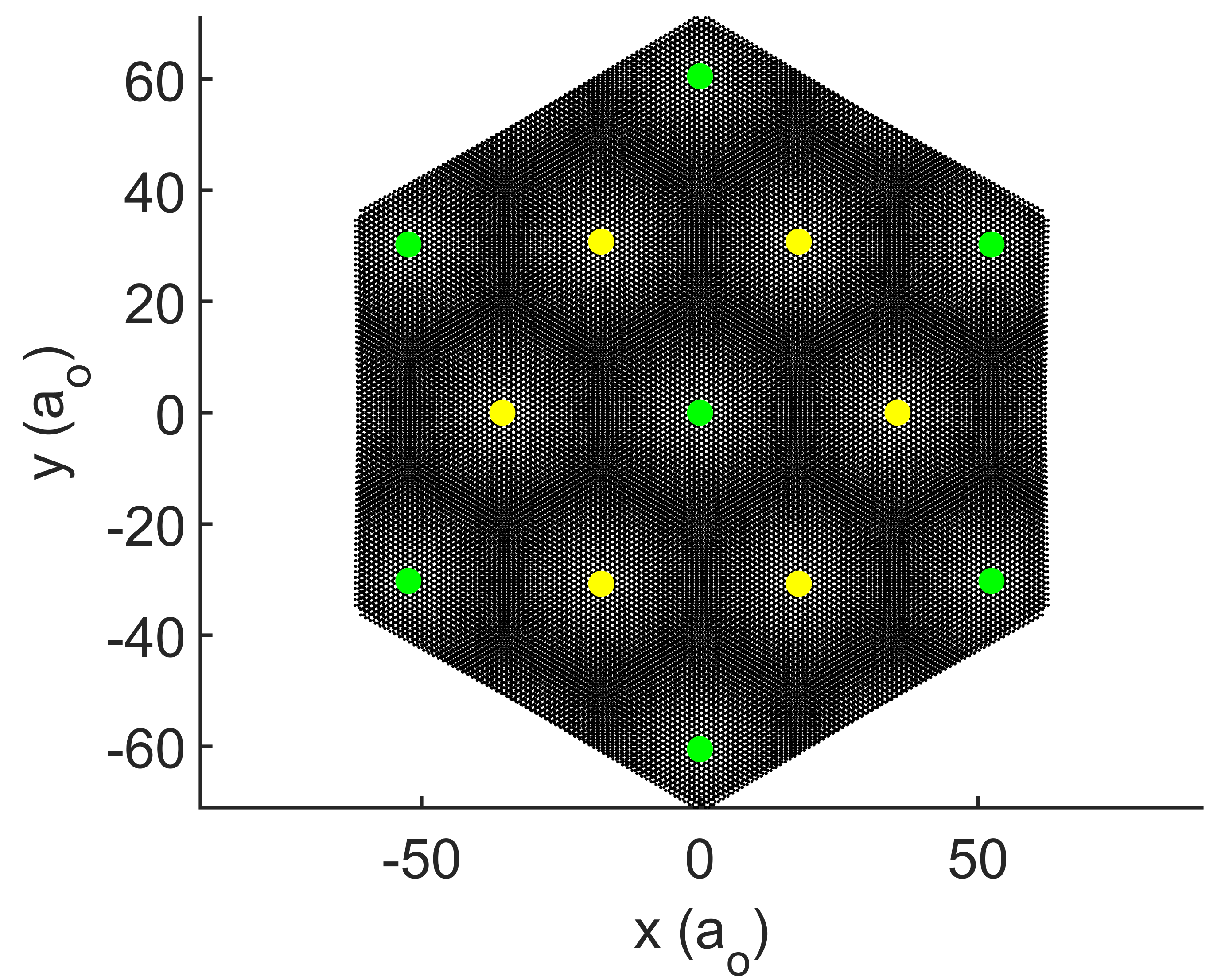}\label{fig6b}}\\
		\caption{(a) Simulated moire pattern of TBG for $\theta_c=1.05^\circ$, $L_c=54.562 a_o$ and $\delta_c=1 a_o$  (b) Simulated moire pattern of TBG for $\theta_c=1.64^\circ$, $L_c=60.506 a_o$ and $\delta_c=\sqrt{3} a_o$; green points are actual moire lattice points and yellow points are  pseudo moire lattice points}	
	\end{figure}
	
	$\bullet$ When we take a closer look on apparent moire period of simulated moire patters of TBG, we find a pattern. For those commensurate twist angles which have the value of corresponding minimum commensurate displacement ($\delta_c$) equal to $a_o$; the apparent moire period of simulated moire patterns (see figure \ref{fig6a}) appears to be same as ($L_c$) the value of moire period supplied in the code; let such moire patterns be named as normal commensurate moire patterns of TBG. For those commensurate twist angles which have the value of corresponding minimum commensurate displacement ($\delta_c$) other than $a_o$; the apparent moire period of simulated moire patterns (see figure \ref{fig6b}) appear quite different from ($L_c$) the value of moire period supplied in the code, let such moire patterns be named as anomalous commensurate moire patterns of TBG.  \\
	
	$\bullet$ To investigate the simulated moire patterns thoroughly we decide to find the value of moire period in simulated moire patterns of TBG. To find the value of moire period in simulated moire pattern we find the locations of actual moire lattice points first. Actual moire lattice points of moire pattern will be those positions where lattice points of $A_1$ and $A_2$ sub-lattices coincide perfectly. To find the positions of actual moire lattice points in simulated TBG the planar distances among various lattice points of $A_1$ and $A_2$ sub-lattices are computed, and the positions of actual moire lattice points are found by locating points where the planar distance between lattice points of $A_1$ and $A_2$ sub-lattices becomes exactly zero.\\
	
	$\bullet$ In normal moire patterns of TBG each AA-type looking region has one actual moire lattice point in the center. In figures of simulated moire patterns of TBG, actual moire lattice points are depicted by green dots. In anomalous moire patterns of TBG, the actual moire lattice points are found to be situated in  the center of some AA-type looking regions while remaining AA-type looking regions  do not have any actual moire lattice point. Those AA-type looking regions,  which do not have any actual moire lattice point; those also seem similar to those AA-type looking regions which have  actual moire lattice points. To investigate deeply we try to locate the points where the planar distance between lattice points of $A_1$ and $A_2$ sub-lattices of two graphene layers becomes very close to zero but not exactly zero. Many such points are found in the center of  each AA-type looking regions of simulated moire pattern of TBG. For those AA-type looking regions  which do not has any actual moire lattice point; we choose one of those points, which has minimum value of planar distance between lattice points of $A_1$ and $A_2$ sub-lattices of two graphene layers, let such points be named as pseudo moire lattice points. In figures of simulated moire patterns of TBG, pseudo moire lattice points are depicted by yellow dots. As the value of $\delta_c$ increases, Number of pseudo moire lattice points increases. Number of pseudo moire lattice points found in a moire unit cell of TBG is found to be equal to $\delta_c^2-1$.\\  
	
   $\bullet$ Actual moire lattice points of simulated moire patterns of TBG appear to be arranged in triangular lattice symmetry. Using the positions of actual moire lattice points, distances among various actual moire lattice points are computed. The distance of any actual moire lattice point from those six actual moire lattice points which appear nearest to it is found each time to be exactly equal to $L_c$ , the value supplied in code. therefore, we conclude that actual moire lattice points form a perfect triangular lattice.\\
   
   $\bullet$ In normal moire patterns of TBG which have the value of corresponding minimum commensurate displacement ($\delta_c$) equal to $a_o$, all the apparent moire lattice points are actual moire lattice points; therefore, they show perfect triangular lattice of moire lattice points. Normal moire patterns of TBG are free from defects such as strain and breaking of rotational symmetry.\\

	\begin{table}[htp]
		\caption{A short list of moire period ($L_c$ in units of $a_o$), corresponding minimum commensurate displacement ($\delta_c$ in units of $a_o$), twist angle ($\theta_c$ in degree), apparent moire period ($L_{app}$ in units of $a_o$), maximum value of apparent moire period ($Max.L_{app}$ in units of $a_o$), minimum value of apparent moire period ($Min.L_{app}$ in units of $a_o$) and apparent strain in moire pattern (in \%) of commensurate TBG }
		\label{Table1}
		\begin{center}
			\begin{tabular}{|c|c|c|c|c|c|c|}
				\hline
				$L_c$  & $\delta_c$ & $\theta_c$ &$L_{app}$&$Max.L_{app}$&$Min.L_{app}$ & Strain  \\
				\hline
				90.0722	&2& 1.2722&45.0361&45.9021&44.1701&3.8457\\
				78.5048	&1.7320 &1.2641&45.3248&45.9021&45.0333&1.9167\\
				45.9021	&1&1.2482&45.9021&45.9021&45.9021&0.0\\
				80.5046	&1.7320 & 1.2327&46.4794&47.6340&45.9021&3.7263\\
				93.5361	&2& 1.2251&46.7680&47.6340&45.9021&3.7033\\
				81.5046	&1.7320& 1.2176&47.0567&47.6340&46.7654&1.8460\\
				47.6340	&1& 1.2028&47.6340&47.6340&47.6340&0.0\\
				83.5045	&1.7320& 1.1884&48.2113&48.5077&47.6340&1.8122\\
				97 &2& 1.1814&48.5&49.3660&47.6340&3.5710\\
				84.5044	&1.7320& 1.1744&48.7887&49.3660&48.4974&1.7802\\
				49.3660	&1&	1.1606&49.3660&49.3660&49.3660&0.0\\
				86.5043	&1.7320& 1.1472&49.9433&51.0979&49.3660&3.4679\\
				87.5043	&1.7320& 1.1341&50.5206&51.0979&50.2295&1.7190\\
				51.0979	&1& 1.1213&51.0979&51.0979&51.0979&0.0\\
				89.5042	&1.7320& 1.1088&51.6753&51.9711&51.0979&1.6898\\
				90.5041	&1.7320& 1.0965&52.2526&52.8299&51.9615&1.6619\\
				52.8299	&1&	1.0845&52.8299&52.8299&52.8299&0.0\\
				92.5040	&1.7320& 1.0728&53.4072&54.5619&52.8299&3.2430\\
				93.5040	&1.7320& 1.0614&53.9846&54.5619&53.6936&1.6085\\
				54.5619	&1&	1.0501&54.5619&54.5619&54.5619&0.0\\
				95.5039	&1.7320& 1.0391&55.1392&55.4346&54.5619&1.5828\\
				96.5039	&1.7320& 1.0284&55.7165&56.2939&54.5619&3.1086\\
				56.2939	&1&	1.0178&56.2939&56.2939&56.2939&0.0\\
				98.5038	&1.7320& 1.0075&56.8712&58.0259&56.2939&3.0455\\
				99.5038	&1.7320& 0.9974&57.4485&58.0259&57.1577&1.5112\\
				58.0258	&1&	0.9874&58.0258&58.0258&58.0258&0.0\\
				\hline	
			\end{tabular}
		\end{center}
	\end{table}

    $\bullet$ In anomalous moire patterns of TBG, the apparent moire lattice points are  found to be blend of both actual moire lattice points and  pseudo moire lattice points. Apparent moire lattice points of anomalous moire patterns of TBG also appear to be arranged in triangular lattice symmetry.  Using the positions of apparent moire lattice points of anomalous moire patterns of TBG, distances among various apparent moire lattice points are computed. In anomalous moire patterns of TBG, the distances of an apparent moire lattices point from those six apparent moire lattice points which appear nearest to it are found to be not always same, these are found to be around  $L_c/\delta_c$. For anomalous moire patterns of TBG we get multiple slightly different values of apparent moire period. From these multiple slightly different values of apparent moire period, maximum ($Max.L_{app}$), minimum ($Min.L_{app}$) and  average ($Avg.L_{app}$) values of apparent moire period are determined. Average value of apparent moire period is found to be very close to $L_c/\delta_c$ , which is named as apparent moire period ($L_{app}$) of moire pattern of TBG.\\
    
    $\bullet$ Apparent moire period of TBG ($L_{app}$) is given by equation \ref{eqn:1} and its variation with twist angle in TBG is shown in figure \ref{fig7a}. $L_{app}$ monotonously decreases with twist angle. A list associating apparent moire period with corresponding commensurate twist angle, minimum commensurate displacement and actual moire period is available on link \cite{300ao_TBG_Lapp_vs_thetac}.\\
    
    \begin{equation}\label{eqn:1}
    	L_{app}= L_c/\delta_c=\frac{1}{2\sin\left( \frac{\theta_c}{2} \right)} \approx Avg.L_{app}
    \end{equation}

    $\bullet$ The apparent strain in anomalous moire pattern of TBG is calculated using following formula:
    (Strain in \%)= $100\left(\frac{Max.L_{app}-Min.L_{app}}{L_{app}}\right)$. Table \ref{Table1} present a small list of computed apparent strain in moire pattern of commensurate TBG for some commensurate twist angles near magic angle. As the value of $\delta_c$ increases, deviation in apparent moire period $\left( Max.L_{app}-Min.L_{app}\right)$ and strain also increase. Therefore, as the value of $\delta_c$ increases randomness in distribution of apparent moire lattice points increases.
     
	$\bullet$ Presence of similar looking pseudo moire lattice points along with actual moire lattice points is the intrinsic reason behind appearance of strain and rotational symmetry breaking in anomalous moire patterns of TBG. Therefore, experimentally observed strain and rotational symmetry breaking in moire patterns of TBG may be due to misunderstanding pseudo moire lattice points as actual moire lattice points.\\

	$\bullet$ For those twist angles which correspond to large value of apparent moire period along with small value of minimum commensurate displacement ($L_{app}>>\delta_c\neq a_o$); a small number of pseudo moire lattice points  get distributed  in large area of moire pattern with small randomness, therefore, the periodicity of moire pattern seems to persist. Therefore those moire patterns of TBG which correspond to large value of apparent moire period along with small value of minimum commensurate displacement ($L_{app}>>\delta_c \neq a_o $) seem to be triangular lattice of moire lattice points with small strain. \\
	
	$\bullet$ For those twist angles which correspond to small value of apparent moire period along with large value of minimum commensurate displacement ($L_{app}<<\delta_c \neq a_o$); a large number of pseudo moire lattice points get distributed in small area of moire pattern with large randomness, therefore, the periodicity of moire pattern seems to got destroyed. Therefore those moire patterns of TBG which correspond to small value of apparent moire period along with large value of minimum commensurate displacement ($L_{app}<<\delta_c \neq a_o$) seem to be incommensurate..  
	
	\section{corrugation in TBG}\label{sec4}
	In conventional bi-layer graphene distance between two graphene layers is found to be uniform everywhere. In AB-stacked bi-layer graphene this inter-layer distance is found to be $d_{\perp AB}=3.35 $\AA\cite{2013-McCann} and in AA-stacked bi-layer graphene this inter-layer distance is found to be $d_{\perp AA}=3.55 $\AA\cite{2008-Lee}. Unlike conventional bi-layer graphene, inter-layer distance in TBG is  not found to be uniform everywhere. TBG has corrugated structure \cite{2019-Alexander,2014-Kazuyuki,2019-Procolo} which appears as there are 2-dimensional cosine like peaks (crests) in center of each AA-type looking regions of TBG which lower to minimum (troughs) in center of AB-type looking regions. This variation of inter-layer distance in TBG is called as corrugation in TBG. Evolution of corrugation in TBG can be described as follows.\\
	
    In AA-stacked bi-layer graphene the inter-layer distance is maximum ($d_{\perp AA}=3.55  $\AA) but it is least stable bi-layer structure of graphene; on the other hand in AB-stacked bi-layer graphene the inter-layer distance is minimum ($d_{\perp AB}=3.35 $\AA) and it is most stable bi-layer structure of graphene. In AA-stacked bi-layer graphene all the $p_z$ orbitals and all C-C bonds of two graphene layers are perfectly aligned. In AB-stacked bi-layer graphene only half of the $p_z$ orbitals of two graphene layers are aligned and all C-C bonds of two graphene layers are most misaligned among all bi-layer graphene structures. Two graphene layers of AB-stacked bi-layer graphene are attracted towards each other by a force which holds them together, but this attractive force is so weak that graphene layers can slide on one  another easily. From these observations we can infer that partial overlap of $p_z$ orbitals of two graphene layers causes attraction between the graphene layers and repulsion between electron pairs of C-C bonds of two graphene layers causes repulsion between two graphene layers.In AA-stacked bi-layer graphene the repulsion between two graphene layers is maximum, therefore, AA-stacked bi-layer graphene is least stable bi-layer graphene and inter-layer distance in AA-stacked bi-layer graphene is maximum among all bi-layer graphene structures. In AB-stacked bi-layer graphene the repulsion between two graphene layers is minimum, therefore, AB-stacked bi-layer graphene is most stable bi-layer graphene and inter-layer distance in AB-stacked bi-layer graphene is minimum among all bi-layer graphene structures.\\
 
    In TBG, environment of $p_z$ orbitals and C-C bonds is different around each carbon atom in a moire  unit cell and periodic repetition of similar moire unit cells constitute the structure of TBG. In a moire unit cell of TBG, the alignment of C-C bonds of two graphene layers varies from most aligned ,in the center of AA-type looking region,  to least aligned, in the center of AB-type looking region. Therefore, in a moire unit cell of TBG, the repulsion between two graphene layers varies smoothly from maximum, in the center of AA-type looking region, to minimum, in the center of AB-type looking region. This periodically and smoothly varying repulsion between two graphene layers in TBG produces periodically and smoothly varying inter-layer distance or corrugation in TBG, which have maximum inter-layer distance ($Max. d_{\perp TBG}$) in the center of AA-type looking region and minimum inter-layer distance ($Min. d_{\perp TBG}$) in the center of AB-type looking region. \\
     
	Corrugation in TBG would induce strain in network of sigma bonds in graphene . Since the corrugation is caused by weak repulsion between graphene layers which is very weak in comparison to strong C-C sigma bonds, therefore, the magnitude of corrugation ($\Delta d_{\perp TBG}$), which is defined as half of the difference between maximum and minimum inter-layer distance in TBG, would be such that the strain in network of sigma bonds in graphene caused by the corrugation in TBG  remains very small. To minimize the strain in network of sigma bonds in graphene in TBG, height of carbon atoms in graphene plane in TBG would vary so slowly and smoothly that despite corrugation, at local level the atomic structure of individual graphene layers of TBG would not change much. \\
	
	Conventional bi-layer graphene, i.e., TBG corresponding to $0^\circ$ twist angle can be considered as a TBG system consisted of infinitely large AA-stacked bi-layer graphene combined with infinitely large AB-stacked bi-layer graphene. The junction of two regions in such TBG would not be sharp but it will be slowly and smoothly varying transition from one type of region to other type of region, also the span of junction region would be infinitely large. Inter-layer distance in such TBG is maximum in the center of AA-type looking region and minimum in the center of AB-type looking region. In such TBG  the magnitude of corrugation is maximum, but the inter-layer distance changes slowly and smoothly over infinitely large span, therefore, the strain in sigma bonds network of graphene in such TBG would be almost zero. \\
	
	As the twist angle increases from $0^\circ$, C-C bonds of two layers move towards more aligned structure in AB-type looking region and C-C bonds of two layers move towards less aligned structure in AA-type looking region; therefore, the repulsion between two graphene layers decreases in AA-type looking region while the repulsion between two layers increases in AB-type looking region. Therefore, as the twist angle increases maximum inter-layer distance of TBG decreases and minimum inter-layer distance of TBG increases. Since the moire pattern corresponding to twist angle $\theta$ and $60^\circ-\theta$ are practically same; therefore, as the twist angle increases, maximum interlayer distance of TBG decreases and minimum interlayer distance of TBG increases till the twist angle of $30^\circ$ is reached and after the twist angle of $30^\circ$ the trend of maximum and minimum inter-layer distance is reversed symmetrically. \\
	
	To minimize strain in network of sigma bonds of graphene in TBG; in comparison to apparent moire period the magnitude of corrugation should be very small and its variation with twist angle should be as slow or fast, as the variation of apparent moire period with twist angle. For those twist angles which are very close to $0^\circ$, the apparent moire period will be so large that even with maximum magnitude of corrugation, the strain induced in network of sigma bonds of graphene in TBG would be practically almost zero. Therefore, for those twist angles which are very close to $0^\circ$, the magnitude of corrugation would decrease so slowly that practically it would be same as maximum value of magnitude of corrugation.\\
	
	This type of behavior of inter-layer distance in TBG can be represented mathematically by following formulation:
	
	\begin{equation}\label{eqn:2}
		\fl\Delta d_{\perp TBG}=\frac{\left(\begin{array}{c}
				\alpha_{11}\left(d_{\perp AA}-d_{\perp AB}\right)\exp\left(-\alpha_{12}|\sin\left(3\theta_C\right)|^{\alpha_{13}}\right)\\
				+\alpha_{21}\left(d_{\perp AA}-d_{\perp AB}\right)\exp\left(-\alpha_{22}|\sin\left(3\theta_C\right)|^{\alpha_{23}}\right)\\
				+\alpha_{31}\left(d_{\perp AA}-d_{\perp AB}\right)\exp\left(-\alpha_{32}|\sin\left(3\theta_C\right)|^{\alpha_{33}}\right)\\
				+0.001\left(\alpha_{11}+\alpha_{21}+\alpha_{31}\right)\left(d_{\perp AA}-d_{\perp AB}\right)\left(1-\exp\left(-|\sin\left(3\theta_C\right)|\right)\right)
			\end{array}\right)}{2\left(\alpha_{11}+\alpha_{21}+\alpha_{31}\right)}
	\end{equation}
	
	\begin{equation}\label{eqn:3}
		Avg. d_{\perp TBG}=0.5(d_{\perp AA}+d_{\perp AB})
	\end{equation}
	
	\begin{equation}\label{eqn:4}
		Max. d_{\perp TBG}=Avg. d_{\perp TBG}+\Delta d_{\perp TBG}
	\end{equation}
	
	\begin{equation}\label{eqn:5}
		Min. d_{\perp TBG}=Avg. d_{\perp TBG}-\Delta d_{\perp TBG}
	\end{equation}

\begin{figure} [htp]
	\centering
	\subfloat[]{\includegraphics[width=0.49\linewidth]{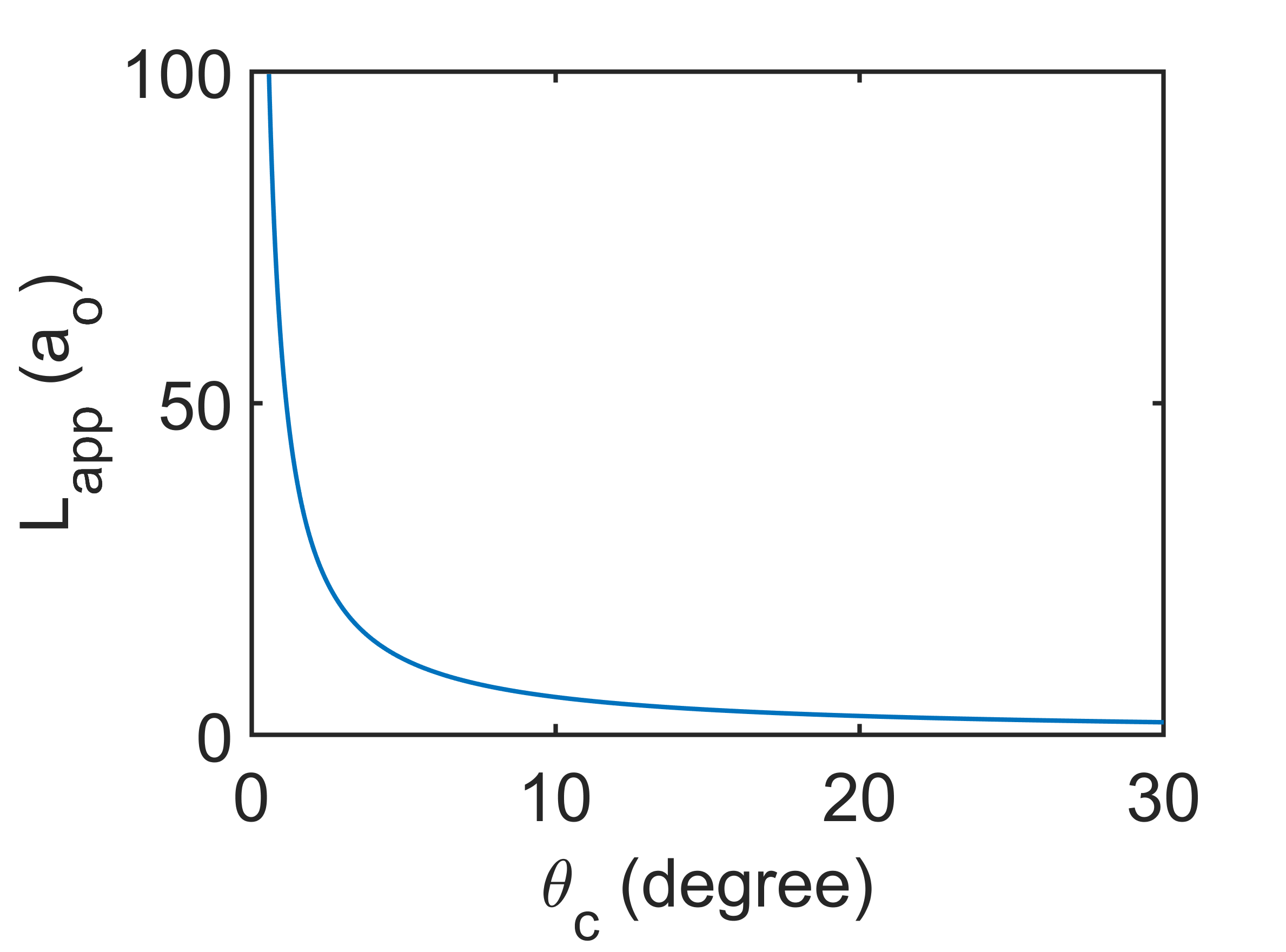}\label{fig7a}}
	\subfloat[]{\includegraphics[width=0.49\linewidth]{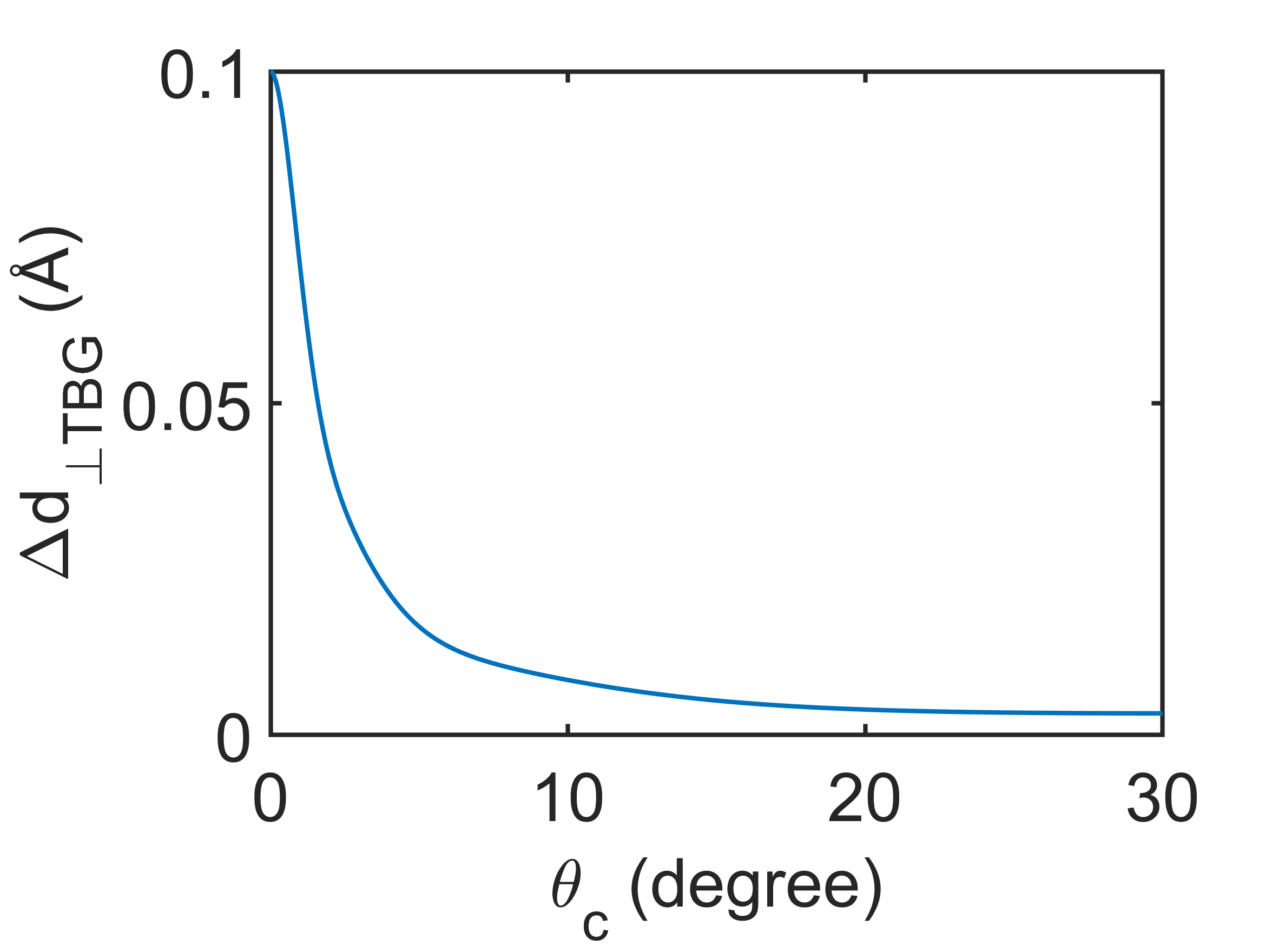}\label{fig7b}}\\
	\subfloat[]{\includegraphics[width=0.49\linewidth]{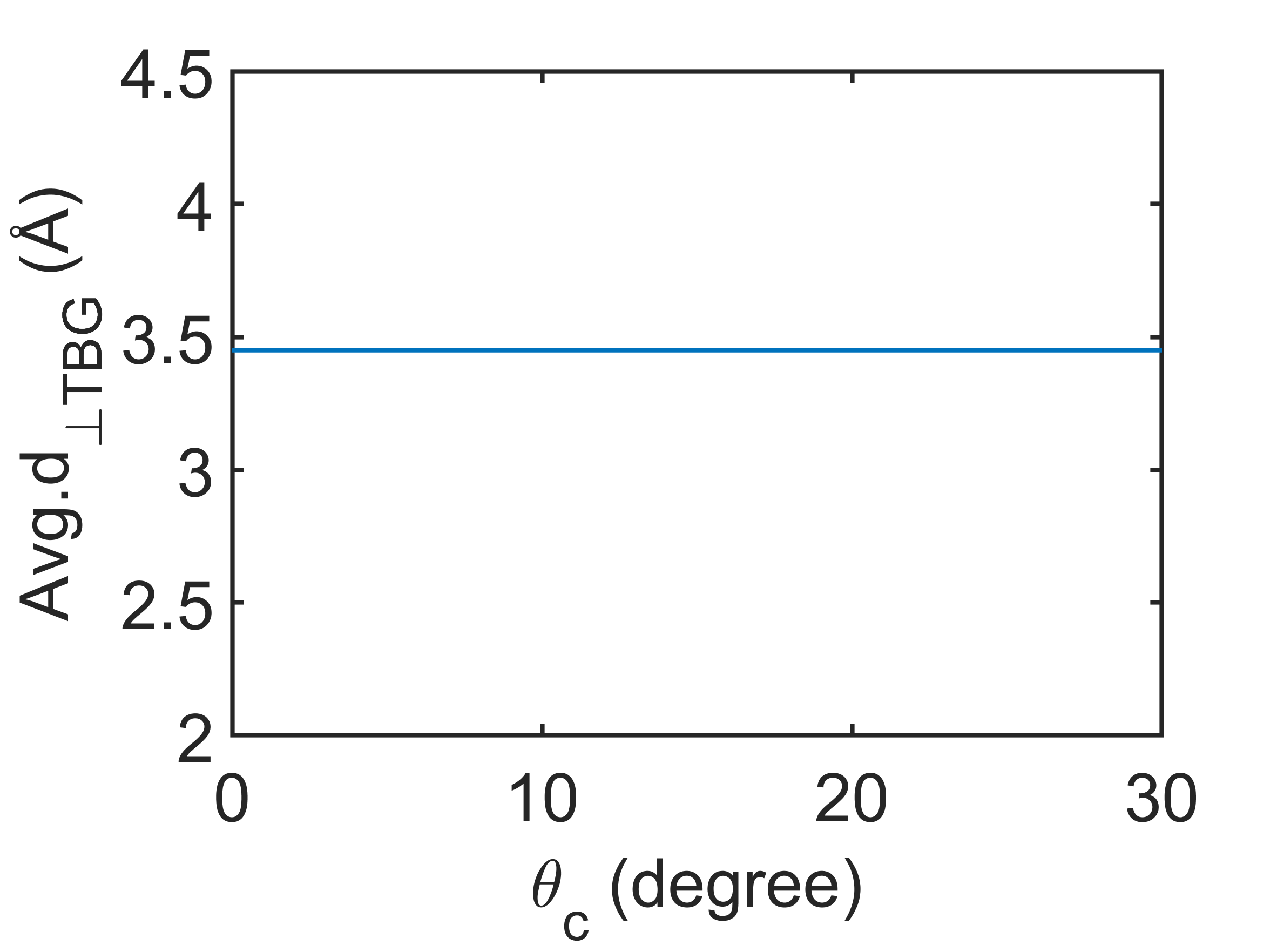}\label{fig7c}}
	\subfloat[]{\includegraphics[width=0.49\linewidth]{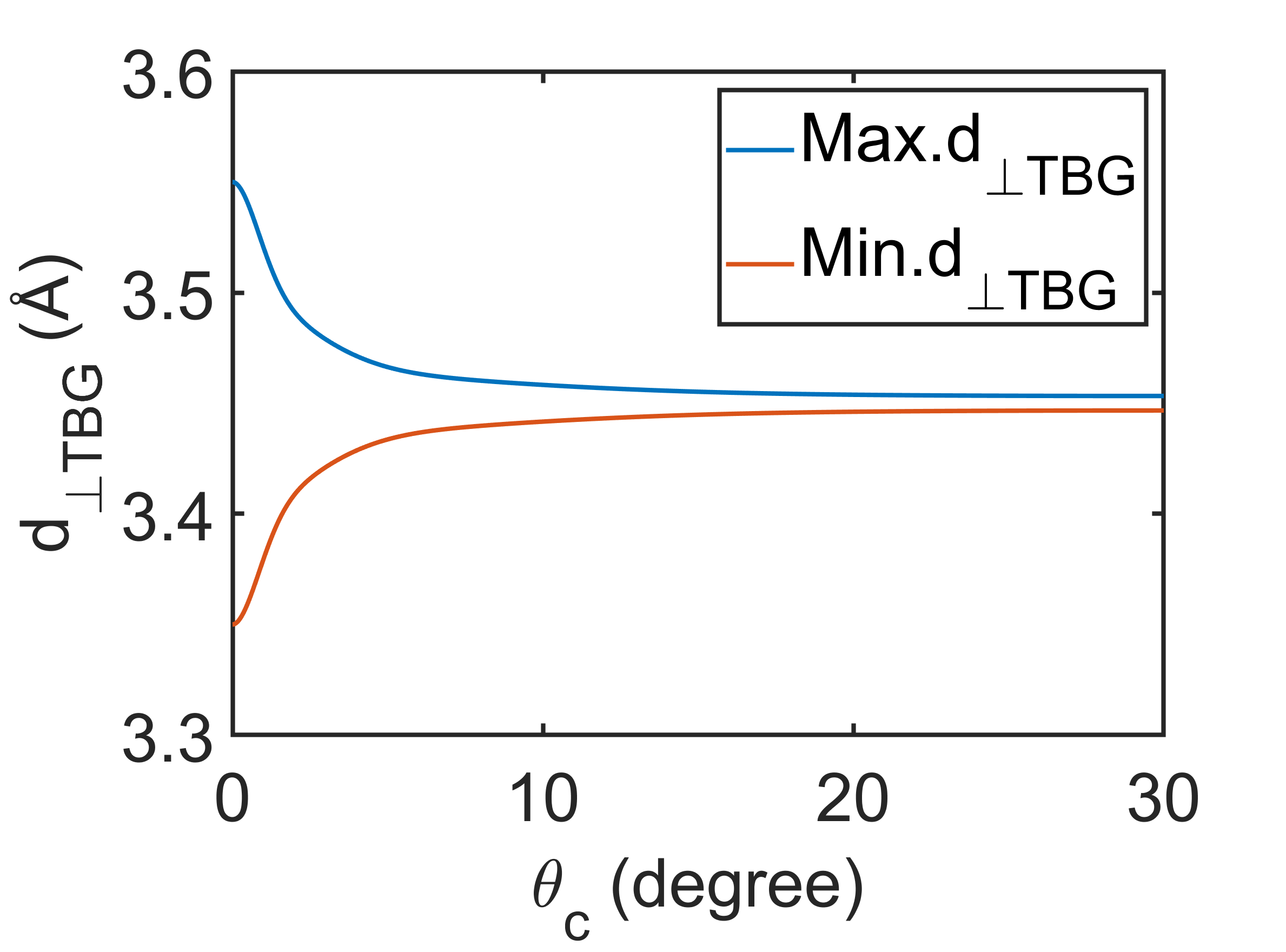}\label{fig7d}}
	\caption{(a) Variation of apparent moire period with twist angle in TBG (b) Variation of magnitude of corrugation with twist angle in TBG, (c) variation of average inter-layer distance with twist angle in TBG (d) Variation of maximum and minimum inter-layer distance with twist angle in TBG}	
\end{figure}
	
	 Magnitude of corrugation is given by equation \ref{eqn:2}. Variation of magnitude of corrugation with twist angle in TBG is shown in figure \ref{fig7b}. Values of parameters used in equation\ref{eqn:2} are $\alpha_{11}=10$, $\alpha_{12}=200$, $\alpha_{13}=1.9$, $\alpha_{21}=7$, $\alpha_{22}=30$, $\alpha_{23}=1.9$, $\alpha_{31}=3$, $\alpha_{32}=3$, $\alpha_{33}=1.9$; which are estimated by attempting to achieve same type of variation of magnitude of corrugation with twist angle as the apparent moire period  varies with twist angle.To achieve this type of variation of magnitude of corrugation with twist angle ; we tried many functions based on our knowledge of mathematical functions; the function presented by equation \ref{eqn:2}, gave best results. Average inter-layer distance ($Avg. d_{\perp TBG}$) is given by equation\ref{eqn:3}, it is same for all twist angles in TBG and its variation with twist angle in TBG is shown in figure \ref{fig7c}. Maximum ($Max. d_{\perp TBG}$) and minimum ($Min. d_{\perp TBG}$) inter-layer distance in TBG are given by equation\ref{eqn:4} and equation\ref{eqn:5} respectively and their variation with twist angle in TBG is shown in figure \ref{fig7d}. Magnitude of corrugation given by equation \ref{eqn:2} produces similar variation of inter-layer distance in TBG with twist angle as presented in the paper associated with the reference \cite{2014-Kazuyuki}. In that paper the authors have obtained the variation of inter-layer distance in TBG with twist angle through vdW-DFT and LDA-DFT calculations, but they did not provide explicit formula for magnitude of corrugation. Although our explanation for corrugation is different from that of vdW-DFT and LDA-DFT calculations but the results show good agreement.

	After determining the magnitude of corrugation in TBG, z-coordinates of carbon atoms in two graphene layers of TBG are determined on the basis of their distance from the center of nearest AA-type looking region. z-coordinates of carbon atoms in two graphene layers of TBG are given by equations \ref{eqn:6a}, \ref{eqn:6b}, \ref{eqn:6c}, \ref{eqn:6d}.
	
	\numparts\begin{eqnarray}	
		\fl	Z_{A_1\left(m\right)}=-0.5\left(Min. d_{\perp TBG}+\left(Max. d_{\perp TBG}-Min. d_{\perp TBG}\right)\left(\cos\left(\frac{\pi}{2a}\left( r^{\parallel}_{A_1\left(m\right),nearest\, A_2}\right)\right)\right)^{\alpha_o}\right) \qquad\label{eqn:6a}\\
		\fl	Z_{B_1\left(m\right)}=-0.5\left(Min. d_{\perp TBG}+\left(Max. d_{\perp TBG}-Min. d_{\perp TBG}\right)\left(\cos\left(\frac{\pi}{2a}\left( r^{\parallel}_{B_1\left(m\right),nearest\, B_2}\right) \right)\right)^{\alpha_o}\right) \qquad\label{eqn:6b}\\
		\fl	Z_{A_2\left(m\right)}=0.5\left(Min. d_{\perp TBG}+\left(Max. d_{\perp TBG}-Min. d_{\perp TBG}\right)\left(\cos\left(\frac{\pi}{2a}\left( r^{\parallel}_{A_2\left(m\right),nearest\, A_1}\right) \right)\right)^{\alpha_o}\right) \qquad\label{eqn:6c}\\
		\fl	Z_{B_2\left(m\right)}=0.5\left(Min. d_{\perp TBG}+\left(Max. d_{\perp TBG}-Min. d_{\perp TBG}\right)\left(\cos\left(\frac{\pi}{2a}\left( r^{\parallel}_{B_2\left(m\right),nearest\, B_1}\right) \right)\right)^{\alpha_o}\right) \qquad\label{eqn:6d} 		
	\end{eqnarray}\endnumparts
	
	\begin{figure} [htp]
		\centering
		\subfloat[]{\includegraphics[width=0.49\linewidth]{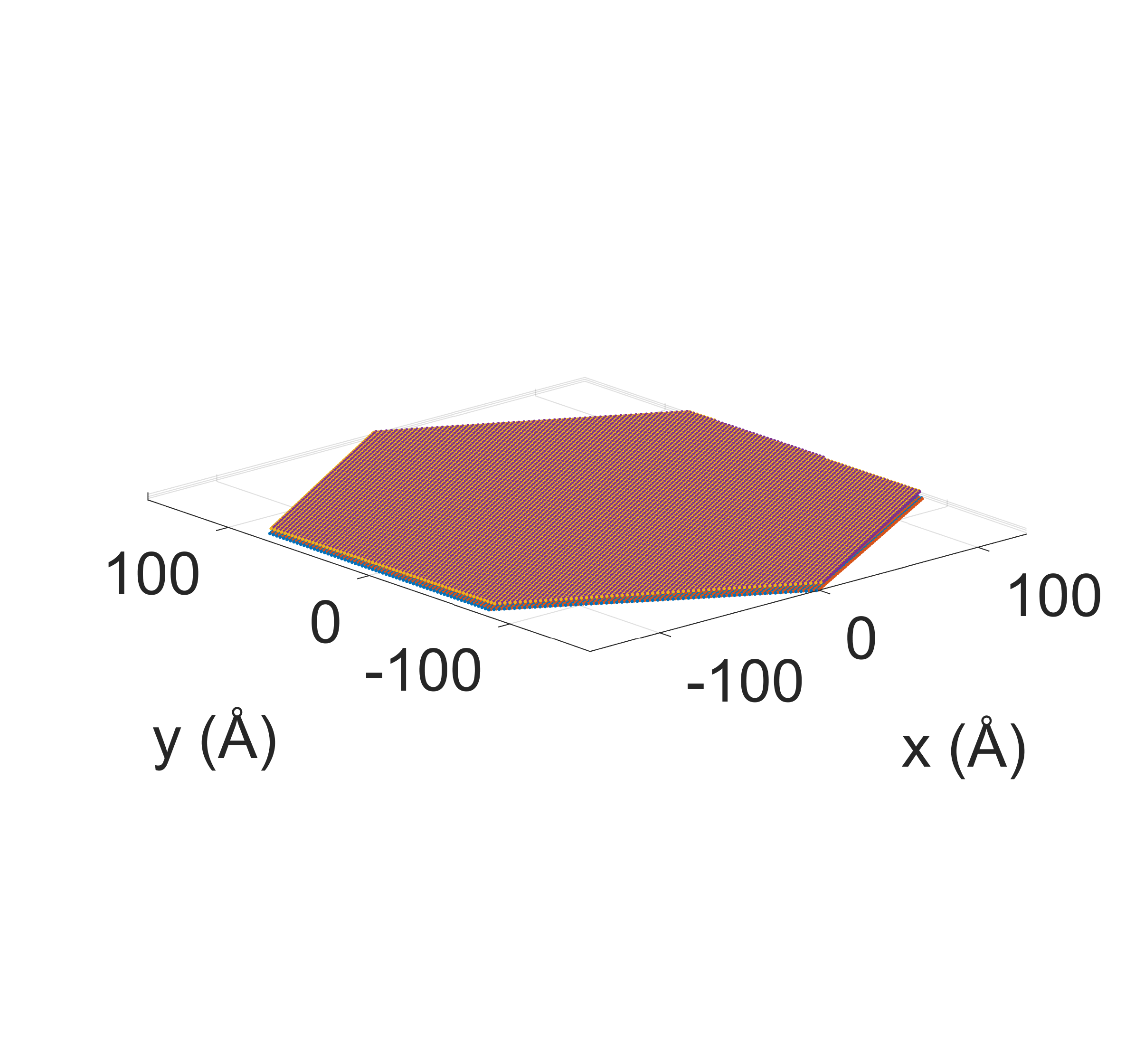}\label{fig8a}}
		\subfloat[]{\includegraphics[width=0.49\linewidth]{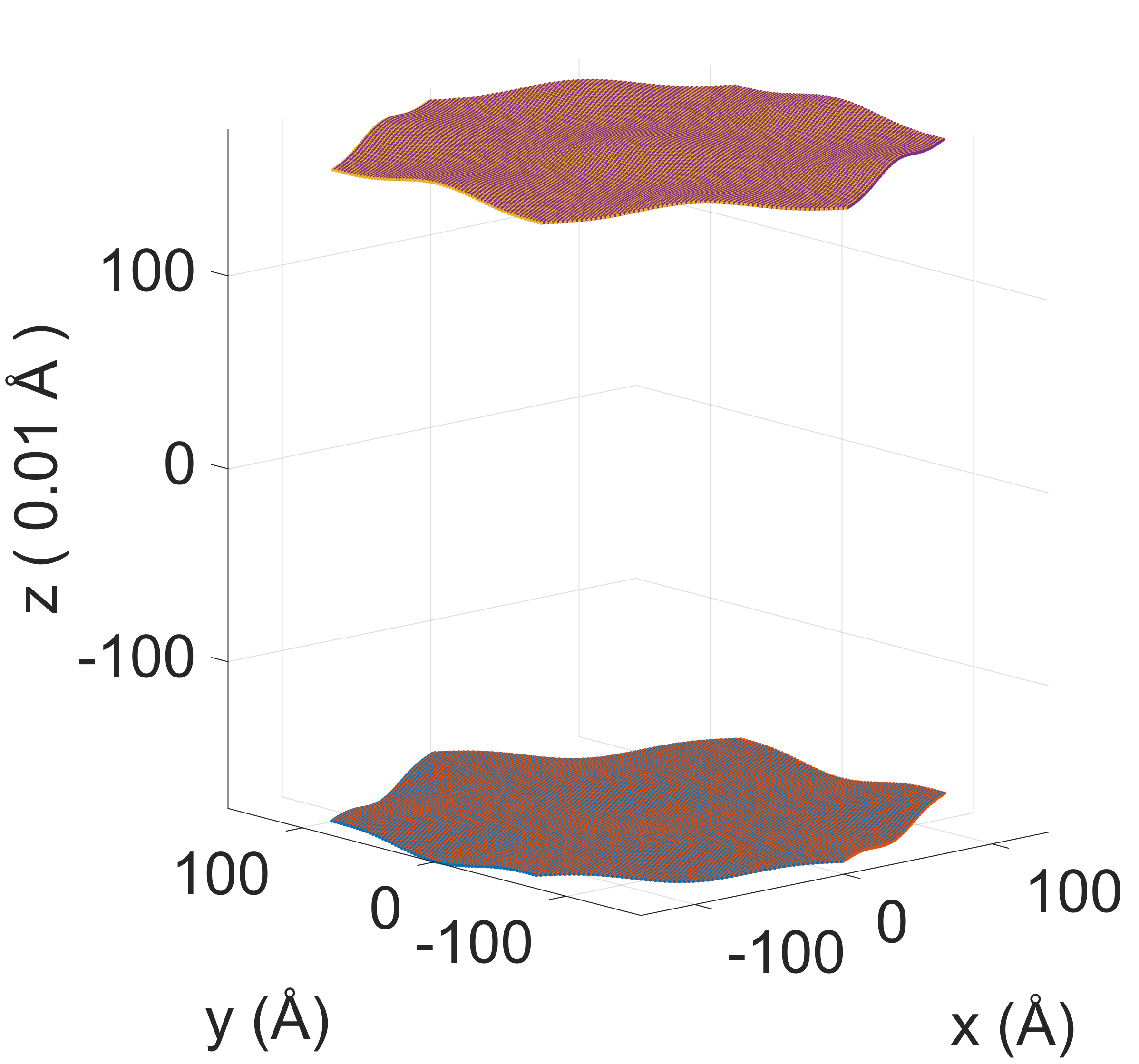}\label{fig8b}}\\
		\subfloat[]{\includegraphics[width=0.49\linewidth]{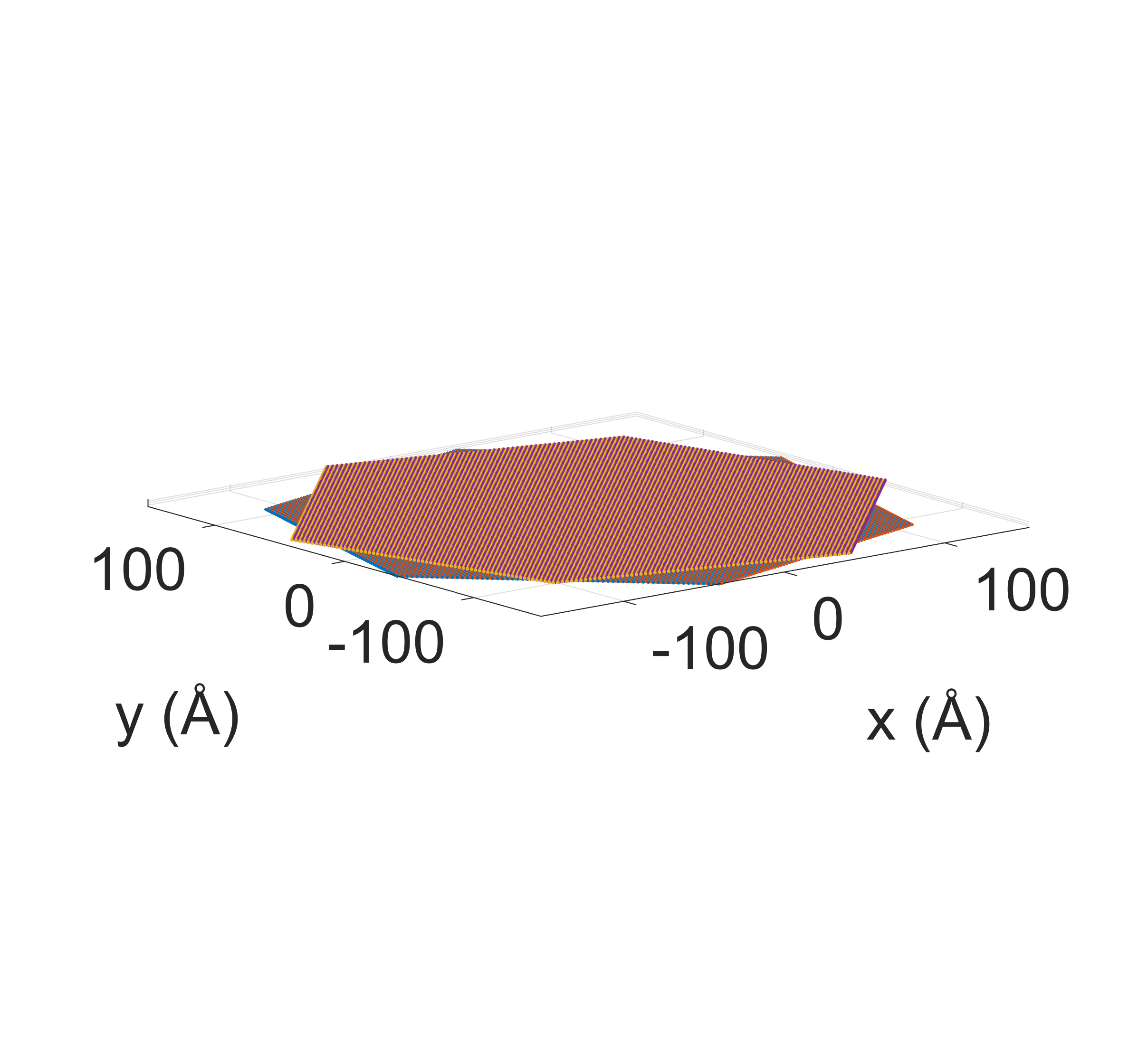}\label{fig8c}}
		\subfloat[]{\includegraphics[width=0.49\linewidth]{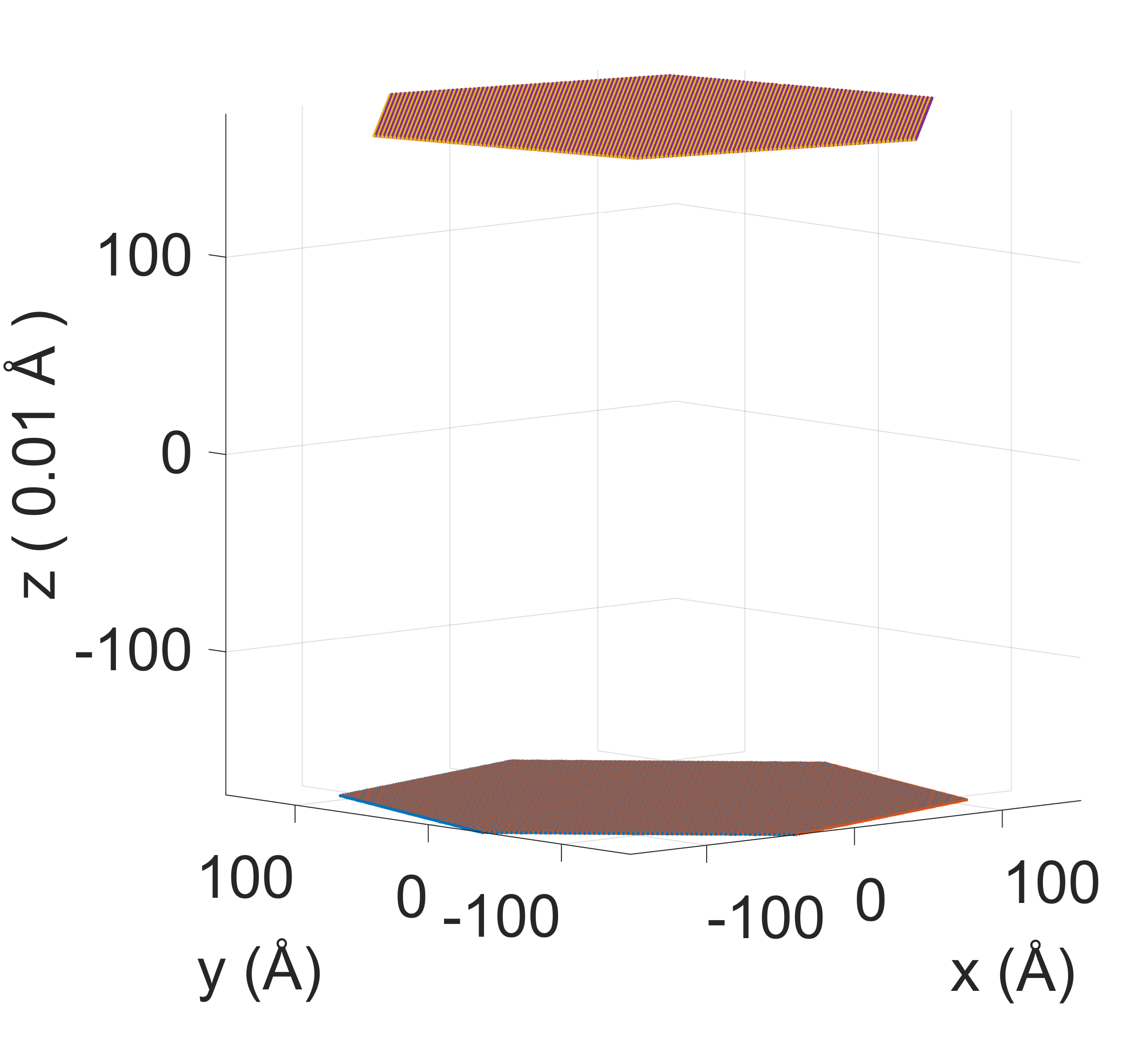}\label{fig8d}}
		\caption{(a),(b) Corrugation in TBG for twist angle equal to $1.0501^\circ$ (c), (d) Corrugation in TBG for twist angle equal to $29.9886^\circ$ }\label{fig8}
	\end{figure}
	
	$A_1\left(m\right)$ denotes that carbon atom which is situated at $m^{th}$ lattice point of $A_1$ sub-lattice; $B_1\left(m\right)$ denotes that carbon atom which is situated at $m^{th}$ lattice point of $B_1$ sub-lattice; $A_2\left(m\right)$ denotes that carbon atom which is situated at $m^{th}$ lattice point of $A_2$ sub-lattice; and $B_2\left(m\right)$ denotes that carbon atom which is situated at $m^{th}$ lattice point of $B_2$ sub-lattice. $Z_{A_1\left(m\right)}$, $Z_{B_1\left(m\right)}$, $Z_{A_2\left(m\right)}$, and $Z_{B_2\left(m\right)}$ denote the z-coordinates of  those carbon atoms which are respectively denoted by $A_1\left(m\right)$, $B_1\left(m\right)$, $A_2\left(m\right)$ and $B_2\left(m\right)$. $r^{\parallel}_{A_1\left(m\right),nearest\, A_2} $ denotes the planar distance  (in units of $a$) of the carbon atom denoted by $A_1\left(m\right)$ from that carbon atom of $A_2$ sub-lattice which lies nearest to $A_1\left(m\right)$. $r^{\parallel}_{B_1\left(m\right),nearest\, B_2} $ denotes the planar distance (in units of $a$) of the carbon atom denoted by $B_1\left(m\right)$ from that carbon atom of $B_2$ sub-lattice which lies nearest to  $B_1\left(m\right)$. $r^{\parallel}_{A_2\left(m\right),nearest\, A_1} $ denotes the planar distance (in units of $a$) of the carbon atom denoted by $A_2\left(m\right)$  from that carbon atom of $A_1$ sub-lattice which lies nearest to $A_2\left(m\right)$. $r^{\parallel}_{B_2\left(m\right),nearest\, B_1} $ denotes the planar distance (in units of $a$) of the carbon atom denoted by $B_2\left(m\right)$ from that carbon atom of $B_1$ sub-lattice which lies nearest to $B_2\left(m\right)$. $\alpha_o=3.0$ is a parameter to control smoothness of corrugation in TBG. In the set of equations (\ref{eqn:6a}-\ref{eqn:6d}), “cos” is a fitting function. We tried many functions, but “cos” suited best for corrugation in TBG.

	We modified that computational code, which was written to simulate planar position of carbon atoms, by incorporating z-component along with planar components of position of carbon atoms in that code. Both figure\ref{fig8a} and figure\ref{fig8b} show corrugation in structure of TBG for twist angle equal to $1.0501^\circ$. Both figure\ref{fig8c} and figure\ref{fig8d} show corrugation in structure of TBG for twist angle equal to $29.9886^\circ$. In figure\ref{fig8a} and \ref{fig8c} the scale of all axes is same, and it seems that both graphene layers are flat; which indicate that magnitude of corrugation and inter-layer distance  in TBG are very small in comparison to period of moire pattern. In figure\ref{fig8b} and \ref{fig8d} the scale of planar axes is kept same as in figure \ref{fig8a} and figure \ref{fig8c}; but the scale of z-axis is scaled up 100 times to visualize the corrugation in TBG. In TBG for twist angle equal to $1.0501^\circ$, magnitude of corrugation is $0.0683$\AA, maximum inter-layer distance is ($3.5183$\AA) and minimum inter-layer distance is ($3.3817$\AA). In TBG for twist angle equal to $29.9986^\circ$ magnitude of corrugation is $0.0008$\AA, which is very small; due to very small value of magnitude of corrugation the maximum inter-layer distance ($3.4508$\AA), minimum inter-layer distance ($3.4492$\AA) and average inter-layer distance ($3.45$\AA) are almost same\cite{2021-Yuki}. Research paper associated with reference \cite{2021-Yuki} report that the inter-layer distance in quasi crystal bi-layer graphene (TBG with $30^{\circ}$ twist angle) is around $3.46$ \AA, which is same as the-inter-layer distance produced by our theory. Research paper associated with reference \cite{2014-Kazuyuki} presents some simulated samples of TBG obtained through vdW-DFT and LDA-DFT simulations; although the size of those simulated samples of TBG is small in comparison to our simulated samples but both show smooth structure of graphene layers of TBG with fine resolution. 
	
	\section{Internal configuration of carbon atoms inside the moire unit cell of TBG}\label{sec5}
	
	\begin{figure} [htp]
		\centering
		\subfloat[]{\includegraphics[width=0.49\linewidth]{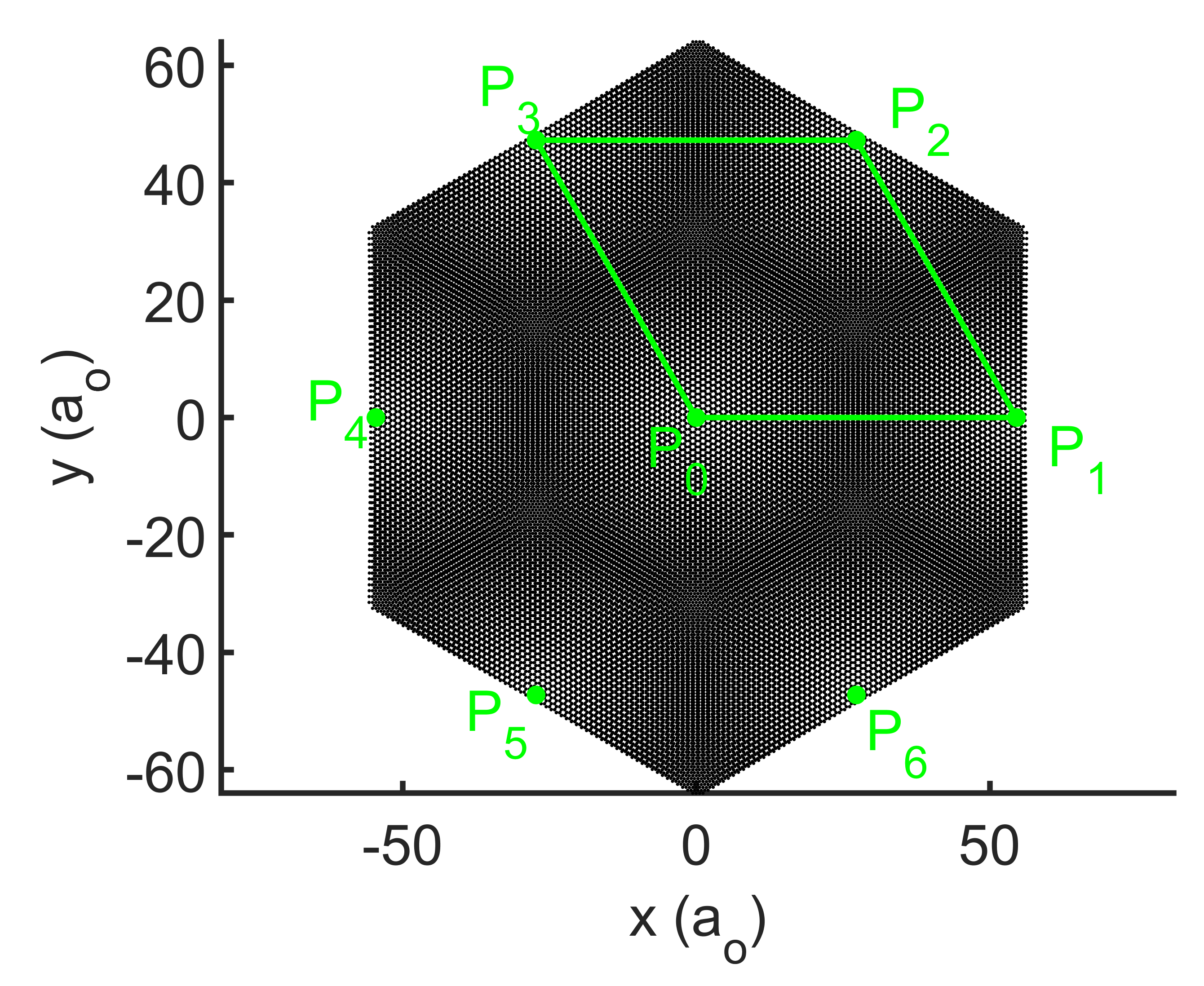}\label{fig9a}}
		\subfloat[]{\includegraphics[width=0.49\linewidth]{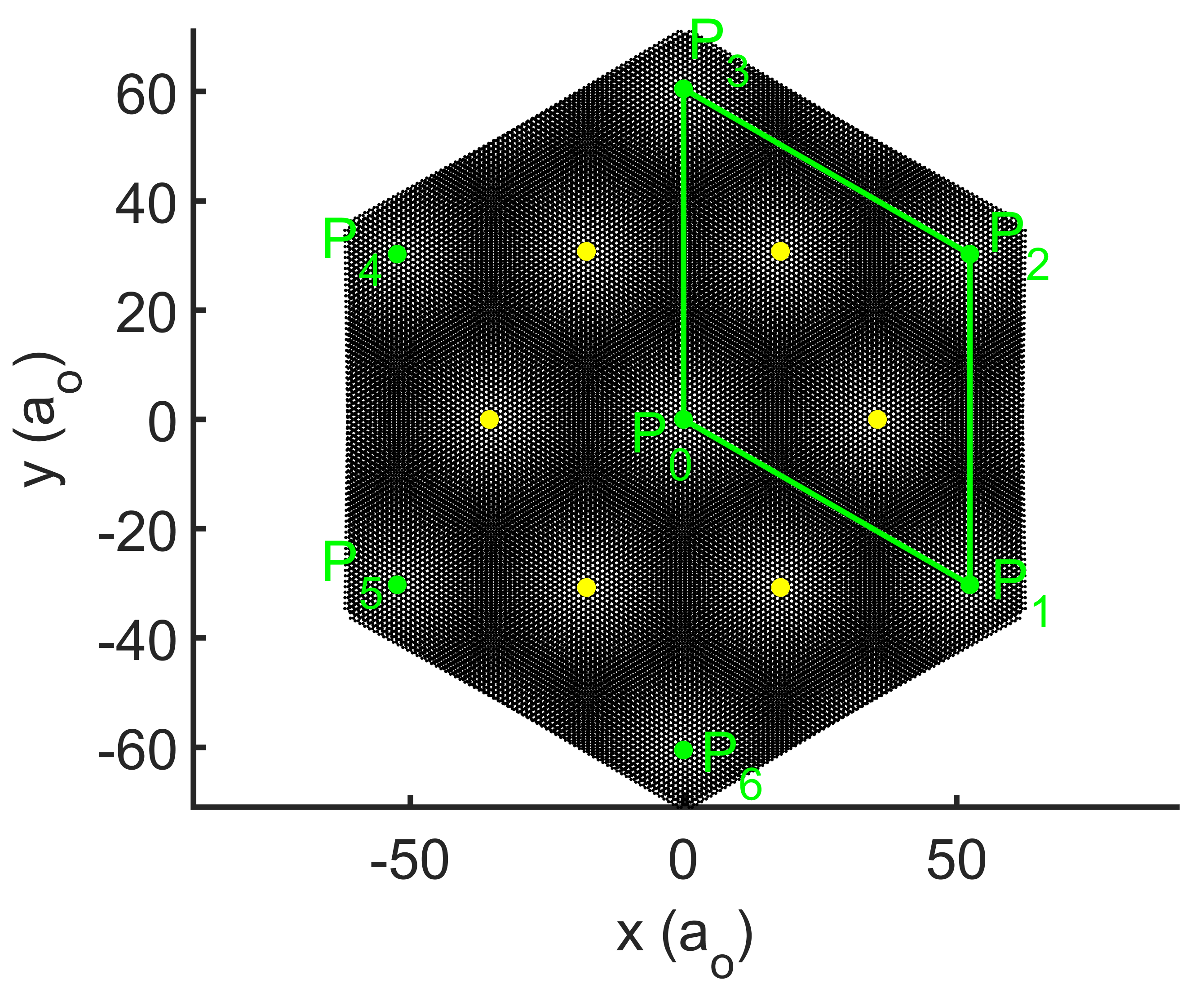}\label{fig9b}}
		\caption{(a)Moire unit cell of moire pattern in TBG for twist angle equal to $1.0501^\circ$ and corresponding minimum commensurate displacement $\delta_c=a_o$(b) Moire unit cell of moire pattern in TBG for twist angle equal to $1.6402^\circ$ and corresponding minimum commensurate displacement $\delta_c=\sqrt{3}a_o$}	
	\end{figure}
	
	While simulating moire pattern in TBG for various commensurate twist angles, we observe a pattern in positions of actual moire lattice points. If we rotate two graphene layers of conventional bi-layer graphene by a relative twist angle $\theta_c$ in such a way that the lower layer gets twisted by angle $\theta_1=-0.5\theta_c$ (in clockwise direction), and upper layer gets twisted by angle $\theta_2=+0.5\theta_c$ (in anticlockwise direction); then the actual moire lattice points, which get generated nearest to origin, are found to be situated at distance $L_c$ from origin, and their positions are found either in directions of midpoint of sides of HEXAGON$^*$ or in directions of corners of HEXAGON$^*$.\\
	
	If the minimum commensurate displacement $\delta_c$ corresponding to twist angle $\theta_c$ is integer multiple of $a_o$ then the positions of actual moire lattice points are found in directions of midpoint of sides of HEXAGON$^*$. If the minimum commensurate displacement $\delta_c$ corresponding to twist angle $\theta_c$ is integer multiple of $\sqrt{3}a_o$ then the positions of actual moire lattice points are found in directions of corners of HEXAGON$^*$.\\
	
	Therefore,  if the minimum commensurate displacement $\delta_c$ corresponding to twist angle $\theta_c$ is integer multiple of $a_o$, then the positions of actual moire lattice points lying  nearest to origin are found to be $P_1\equiv\left(L_c,0.0\right)$, 
	$P_2\equiv\left(\frac{L_c}{2},\frac{\sqrt{3}L_c}{2}\right)$,    $P_3\equiv\left(-\frac{L_c}{2},\frac{\sqrt{3}L_c}{2}\right)$, $P_4\equiv\left(-L_c,0.0\right)$,  
	$P_5\equiv\left(-\frac{L_c}{2},-\frac{\sqrt{3}L_c}{2}\right)$,  
	$P_6\equiv\left(\frac{L_c}{2},-\frac{\sqrt{3}L_c}{2}\right)$ , as shown in figure \ref{fig9a}.  \\
	
	If the minimum commensurate displacement $\delta_c$ corresponding to twist angle $\theta_c$ is integer multiple of $\sqrt{3}a_o$, then the positions of actual moire lattice points lying  nearest to origin are found to be $P_1\equiv\left(\frac{\sqrt{3}L_c}{2},-\frac{L_c}{2}\right)$, 
	$P_2\equiv\left(\frac{\sqrt{3}L_c}{2},\frac{L_c}{2}\right)$, 
	$P_3\equiv\left(0,L_c\right)$, $P_4\equiv\left(-\frac{\sqrt{3}L_c}{2},\frac{L_c}{2}\right)$,  $P_5\equiv\left(-\frac{\sqrt{3}L_c}{2},-\frac{L_c}{2}\right)$, 
	$P_6\equiv\left(0,-L_c\right)$, as shown in figure \ref{fig9b}.\\
	
	$P_0\equiv\left(0.0,0.0\right)$ is actual moire lattice point situated at origin. Region enclosed by parallelogram $P_0P_1P_2P_3$ can be considered as moire unit cell of TBG (shown in figure \ref{fig9a} \ref{fig9b}). Position of carbon atoms lying inside the moire unit cell of TBG can be extracted by using concept of polar angle and some other basic concepts of geometry. In polar coordinate system polar angle is given by: $\varphi=\cos^{-1}\frac{x}{\sqrt{x^2+y^2}}$ if $y\geq0$ and $\sqrt{x^2+y^2}\not=0$ or $\varphi=-\cos^{-1}\frac{x}{\sqrt{x^2+y^2}}$ if $y<0$ or $\varphi=$ undefined if $\sqrt{x^2+y^2}=0$. $\varphi$ lies in range $\left(-180^\circ, 180^\circ\right]$. \\
	
	To extract position of carbon atoms lying inside the moire unit cell of TBG, we made a MATLAB code which workout through following steps:\\ 
	$\bullet$ This code first simulates planar position of carbon atoms in a hexagonal crystal of conventional bi-layer graphene. We may choose any one of conventional bi-layer graphene. We chose AA-stacked bi-layer graphene. Side-length of hexagonal crystal of conventional bi-layer graphene is chosen to be $\frac{2}{\sqrt{3}}L_c$, so that it can at least accommodate those six moire lattice points of TBG which will be generated nearest to origin.\\
	 
	$\bullet$ Then planar position of carbon atoms in moire pattern of TBG are obtained after changing the planar position of carbon atoms of AA-stacked bi-layer graphene according to rule of planar rotational transformation. To obtain TBG from AA-stacked bi-layer graphene, we rotate the two graphene layers in opposite directions by half of the twist angle.\\
	
    $\bullet$ Then z-component of position of carbon atoms of TBG are determined according to their planar positions.\\
    
    $\bullet$ Then the polar angle corresponding to position of each carbon atom of TBG is computed.\\
    
    After this step we have got the position and  polar angle associated with that position of carbon atoms lying in a large crystal of TBG. We consider a set $S_1$ which have one element corresponding to each carbon atom of TBG crystal. Each element of set $S_1$ has information of position and polar angle of its corresponding carbon atom. The position of carbon atoms lying inside the moire unit cell of TBG are extracted from set $S_1$ following two paths. \\

	\begin{figure} [htp]
		\centering
		\subfloat[]{\includegraphics[width=0.49\linewidth]{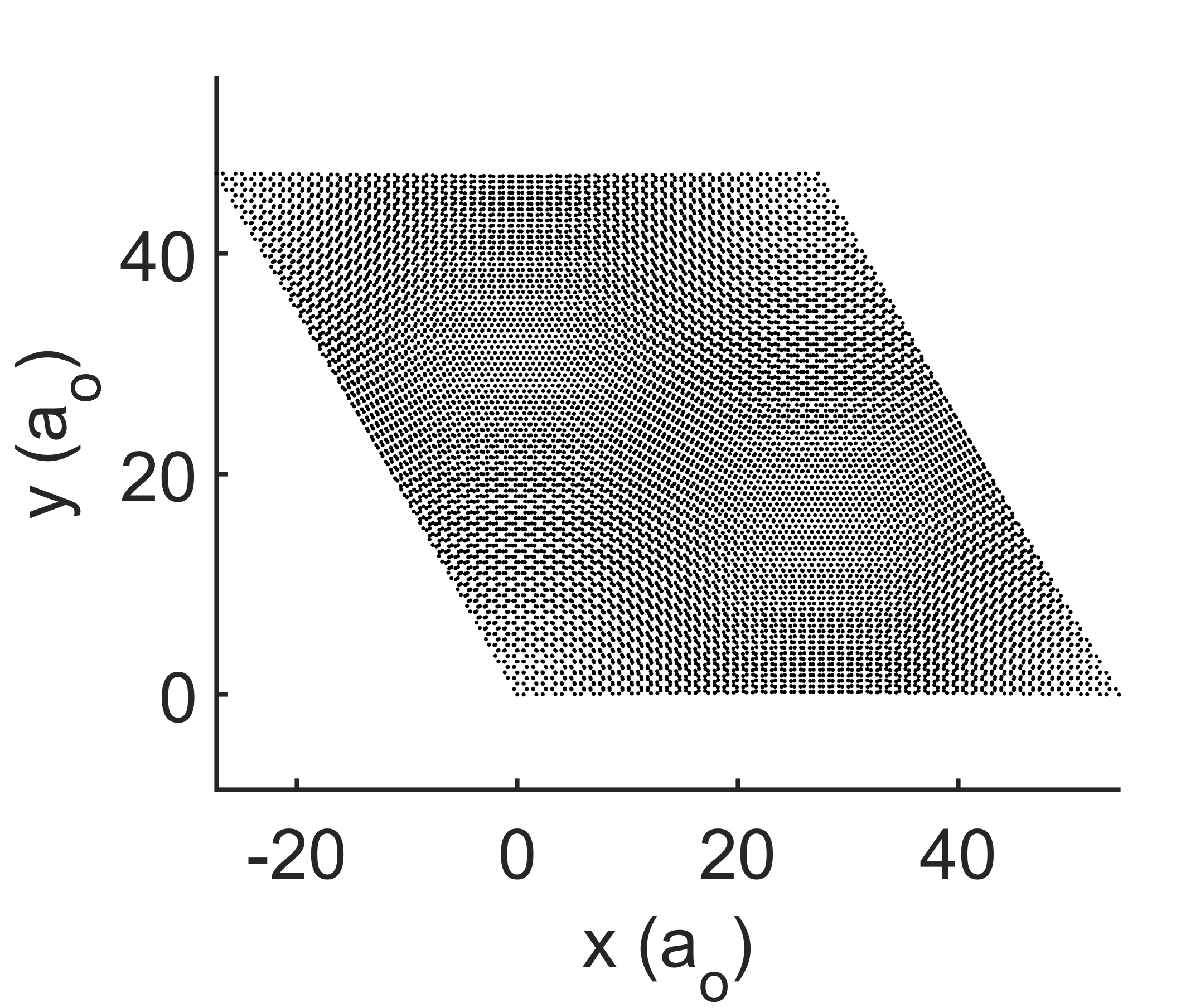}\label{fig10a}}
		\subfloat[]{\includegraphics[width=0.49\linewidth]{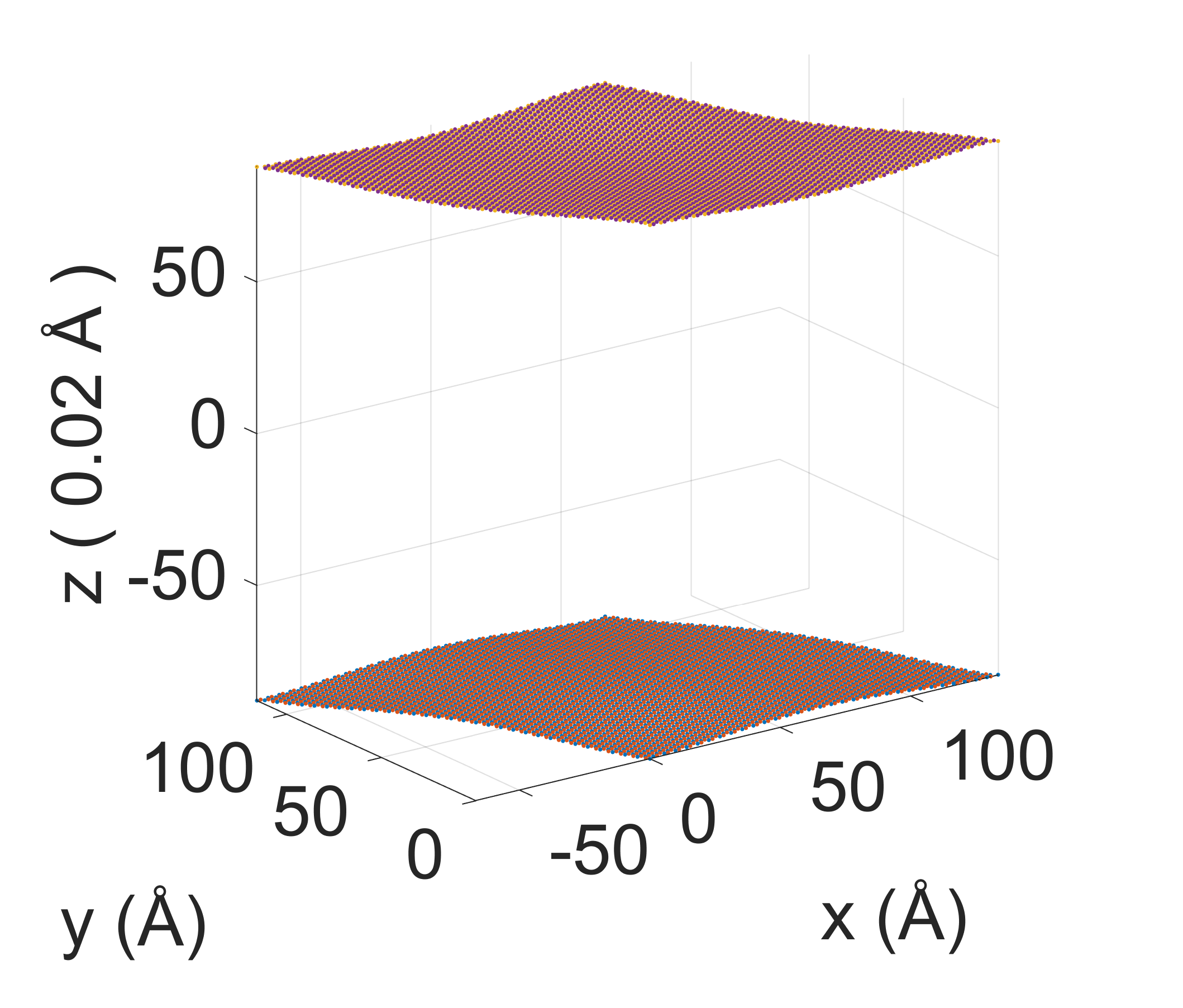}\label{fig10b}}\\
		\subfloat[]{\includegraphics[width=0.49\linewidth]{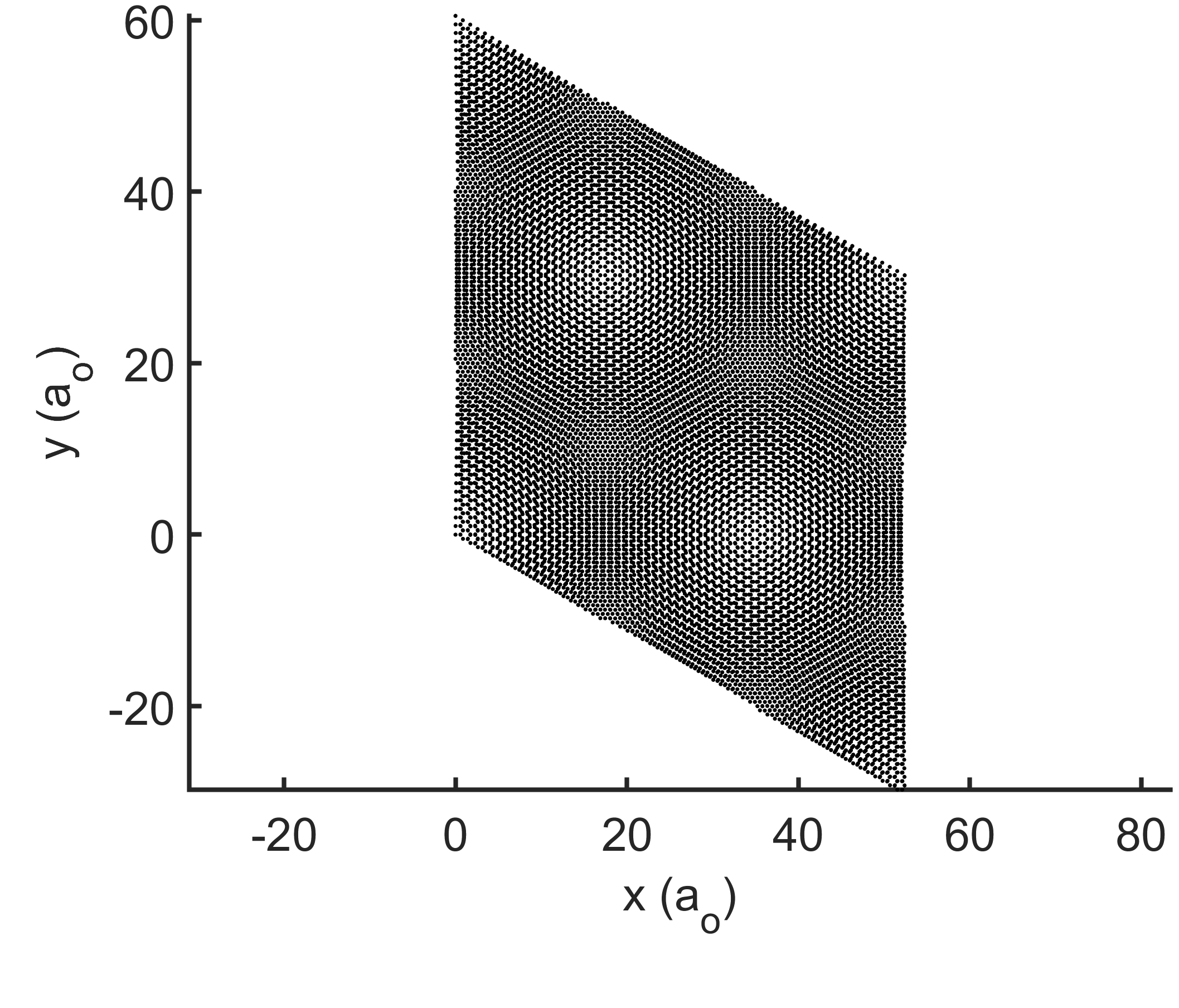}\label{fig10c}}
		\subfloat[]{\includegraphics[width=0.49\linewidth]{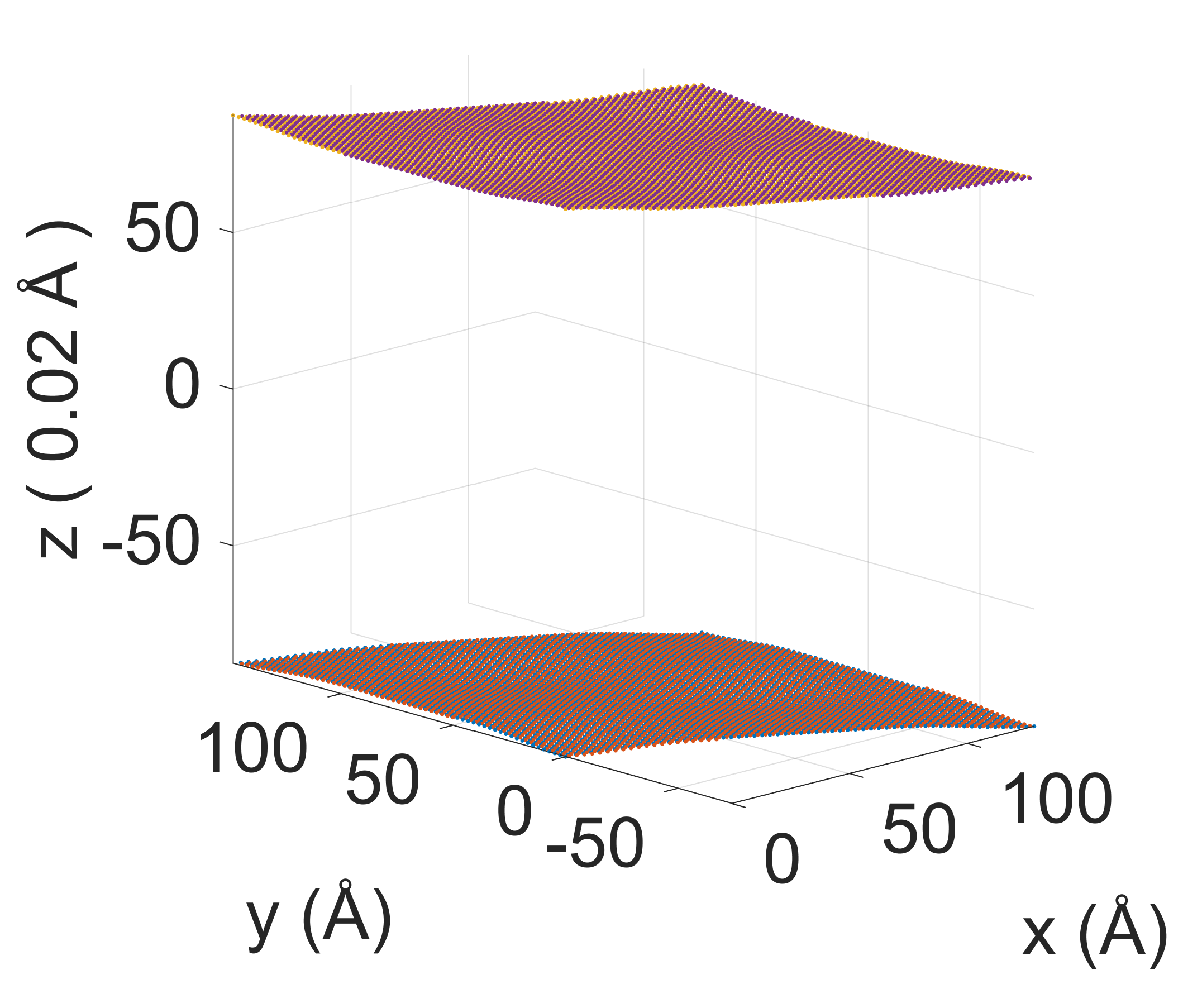}\label{fig10d}}
		\caption{(a), (b) Moire unit cell of moire pattern in TBG for twist angle equal to $1.0501^\circ$ and corresponding minimum commensurate displacement $\delta_c=a_o$(c),(d) Moire unit cell of moire pattern in TBG for twist angle equal to $1.6402^\circ$ and corresponding minimum commensurate displacement $\delta_c=\sqrt{3}a_o$}	
		\label{fig10}
	\end{figure}
	
	Case (1): If the minimum commensurate displacement $\delta_c$ corresponding to twist angle $\theta_c$ is integer multiple of $a_o$ (see moire unit cell in figure \ref{fig9a}): \\
	
	$\bullet$ The value polar angle for position of carbon atoms lying inside the moire unit cell will lie in range $\left[0^\circ, 120^\circ\right]$ . We define a set $S_2$ which is a subset of $S_1$. $S_2$ contains only those elements of $S_1$ which have corresponding value of polar angle lying in range $\left[0^\circ, 120^\circ\right]$.\\ 
	
	$\bullet$ x-coordinate of carbon atoms of moire unit cell will lie in range $\left[-0.5L_c, L_c\right)$ and y-coordinate of carbon atoms of moire unit cell will lie in range $\left[0, L_c\frac{\sqrt{3}}{2}\right)$. We define $S_3$ as a subset of $S_2$ which contains only those elements of $S_2$ that satisfy following condition: x-coordinate of corresponding carbon atom is in range $\left(\frac{1}{2}L_c, L_c\right)$ and y-coordinate of corresponding carbon atom is in range  $\left[0, \sqrt{3}\left(L_c-x\right)\right)$. We define $S_4$ as a subset of $S_2$ which contains only those elements of $S_2$ that satisfy following condition: x-coordinate of corresponding carbon atom is in range $\left[-\frac{1}{2}L_c, \frac{1}{2}L_c\right]$ and y-coordinate of corresponding carbon atom is in range  $\left[0, \frac{\sqrt{3}}{2}L_c\right)$. Moire unit will be union of $S_3$ and $S_4$. \\

	Case:(2) If the minimum commensurate displacement $\delta_c$ corresponding to twist angle $\theta_c$ is integer multiple of $\sqrt{3}a_o$ (see moire unit cell in figure \ref{fig9b}):\\ 
	
	$\bullet$ The value polar angle for position of carbon atoms lying inside the moire unit cell will lie in range $\left[-30^\circ, 90^\circ\right]$. We define a set $S_2$ which is a subset of $S_1$. $S_2$ contains only those elements of $S_1$ which have corresponding value of polar angle lying in range $\left[-30^\circ, 90^\circ\right]$.\\
	
	$\bullet$ x-coordinate of carbon atoms of moire unit cell will lie in range $\left[0, \frac{\sqrt{3}}{2}L_c\right)$ and y-coordinate of carbon atoms of moire unit cell will lie in range $\left[-\frac{x}{\sqrt{3}}, L_c-\frac{x}{\sqrt{3}}\right)$. Moire unit cell will be a subset of $S_2$ which contains only those elements of $S_2$ that satisfy following condition: x-coordinate of corresponding carbon atoms is in range $\left[0, \frac{\sqrt{3}}{2}L_c\right)$ and corresponding y-coordinate of corresponding carbon atoms is in range $\left[-\frac{x}{\sqrt{3}}, L_c-\frac{x}{\sqrt{3}}\right)$.\\

	Figure\ref{fig10} shows the simulated moire unit cell of TBG corresponding to twist angles $1.0501^\circ$ \ref{fig10a}\ref{fig10b} and $1.6402^\circ$\ref{fig10c} \ref{fig10d}. Figure \ref{fig10a} shows the 2D moire  Moire unit cell of TBG for $1.0501^\circ$ twist angle. Figure \ref{fig10b} shows the 3D moire  Moire unit cell of TBG for $1.0501^\circ$ twist angle. Figure \ref{fig10c} shows the 2D moire  Moire unit cell of TBG for $1.6402^\circ$ twist angle. Figure \ref{fig10d} shows the 3D moire  Moire unit cell of TBG for $1.6402^\circ$ twist angle.
	Moire unit cell of TBG corresponding to twist angle $1.0501^\circ$ contains 2977 lattice points of each of $A_1$, $B_1$, $A_2$ and $B_2$ sub-lattices. Moire unit cell of TBG corresponding to twist angle $1.6402^\circ$ contains 3661 lattice points of each of $A_1$, $B_1$, $A_2$ and $B_2$ sub-lattices.
	
	\section{Conclusion}\label{sec6}
	
	$\bullet$ All twist angles are commensurate twist angles which produce perfectly periodic triangular lattice of actual moire lattice points in moire patterns of TBG.\\
	$\bullet$ For characterization of moire pattern of TBG, corresponding minimum commensurate displacement ($\delta_c$) is also important along with commensurate twist angle ($\theta_c$). $\delta_c$ can be either integer multiple of $a_o$ or integer multiple of $\sqrt{3}a_o$.\\
	$\bullet$ If $\delta_c=a_o$, then moire pattern of TBG possess only actual moire lattice points arranged on perfect triangular lattice. \\
	$\bullet$ If $\delta_c\neq a_o$, then moire pattern of TBG possess pseudo moire lattice points along with actual moire lattice points. \\
	$\bullet$ Presence of similar looking pseudo moire lattice points along with actual moire lattice points is the intrinsic reason behind appearance of strain and rotational symmetry breaking in anomalous moire patterns of TBG. Therefore, experimentally observed strain and rotational symmetry breaking in moire patterns of TBG may be due to misunderstanding pseudo moire lattice points as actual moire lattice points.\\
	$\bullet$ For those twist angles which are linked to large value of apparent moire period and small value of minimum commensurate displacement ($L_{app}>>\delta_c\neq a_o$); a small number of pseudo moire lattice points get distributed with small randomness in large area of moire pattern, therefore, the periodicity of moire pattern seems to persist. Therefore those moire patterns of TBG which correspond to large value of apparent moire period and small value of minimum commensurate displacement ($L_{app}>>\delta_c \neq a_o $), seem to be triangular lattice of moire lattice points with small strain. \\
	$\bullet$ For those twist angles which are linked to small value of apparent moire period and large value of minimum commensurate displacement ($L_{app}<<\delta_c \neq a_o$); a large number of pseudo moire lattice points get distributed with large randomness in small area of moire pattern, therefore, the periodicity of moire pattern seems to got destroyed. Therefore those moire patterns of TBG which correspond to small value of apparent moire period and large value of minimum commensurate displacement ($L_{app}<<\delta_c \neq a_o$), seem to be incommensurate.\\
	$\bullet$ There exists corrugation in structure of TBG which is caused by varying value of repulsion between two graphene layers at different points of TBG. This repulsion exists among C-C bonds of two graphene layers in TBG. The repulsion between two graphene layers at different points of TBG varies with varying environment of C-C bond of two graphene layers in TBG. Mathematical model to represent corrugation in TBG, presented in this paper, is good enough to represent corrugation of actual TBG samples. \\
	$\bullet$ There exists a pattern in position of actual moire lattice points in moire pattern of TBG. Using some basic concepts of geometry and computational techniques, it is possible to write a very efficient computational code \cite{AA_to_TBG_unitcell} to simulate internal configuration of carbon atoms inside a moire unit cell of TBG.\\
   $\bullet$ Using the lattice structure of TBG presented in this article, we have shown that \cite{2023-Veerpal_conf1} just environment dependent inter-layer coupling is enough to produce flat band and van hove singularities in TBG.
	
	\ack
	We acknowledge MHRD India for providing research fellowship and Indian Institute of Technology, Roorkee for providing research facilities.
	\section*{References}
	\bibliographystyle{iopart-num}
	\bibliography{bibfile}
	
\end{document}